\documentclass[%
 reprint,
 amsmath,amssymb,
 aps,
prx,
]{revtex4-1}

\usepackage{graphicx}
\usepackage{dcolumn}
\usepackage{bm}

\begin{document}


\title{Non-existence theorems and solutions of the Wu-Yang monopole equation}

\author{Frederik N{\o}rfjand$^1$}
\author{Nikolaj Thomas Zinner$^{1,2}$}%
 
\affiliation{\mbox{$^1$Department of Physics and Astronomy, Aarhus University, Ny Munkegade 120, 8000 Aarhus C, Denmark} \\
\mbox{$^2$Aarhus Institute of Advanced Studies, Aarhus University, H{\o}egh-Guldbergs Gade 6B, 8000 Aarhus C, Denmark}}

\date{\today}

\begin{abstract}
We consider the dynamics of the classical $SU(2)$ Wu-Yang monopole problem and show a set of new directions for its analysis starting from a variational setting. This allows us to give a new interpretation of the monopole charge as a string in a centrifugal potential. The field equations are solved using both standard power series, as well as a new Pad{\'e} approach. Furthermore, we discuss non-existence, and give a proof that rule out even singular non-trivial smooth finite energy solutions. Due to its non-vanishing chromomagnetic fields, the monopole can be interpreted as a collection of magnetic dipoles. Along the same lines, a non-smooth glue-ball solution is discussed that may serve as a toy model of the proton or a radiation-dominated cosmological expansion. Finally, we discuss generalizations of the Wu-Yang monopole.
\end{abstract}

\pacs{Valid PACS appear here}
\maketitle


\section{Introduction}
Magnetic monopoles have been an important subject in physics for almost a century now. Still to this date magnetic monopoles are of much interest to physicists. The story started in 1931 when Dirac proposed the existence of a monopole with magnetic charge \cite{dirac1931quantised}. This was motivated by his result, that the existence of a magnetic monopole obeying Maxwell's equations implies quantization of electric charge. Quantization of electric charge had been a peculiar fact of nature which until Dirac had no explanation. Dirac's monopole is a semi-infinitely long infinitesimally thin solenoid with one free end envisioned as the monopole, behaving as $q_m/r^2$. This is a curious model of a magnetic monopole since it has an infinitely long string attached. Any attempt to find a vector potential for the monopole gives singularities along the string. This singular behaviour is of course necessary to obtain a non-vanishing magnetic charge from Gauss' Law. In 1967 Bais \cite{bais1976configurations} found that by embedding Dirac's monopole into a non-abelian Yang-Mills theory, the solenoid part can be removed by a gauge transformation. This process gives a natural generalization \cite{marciano1978magnetic} of the Wu-Yang monopole \cite{wu1967some}, the latter which is the monopole of interest in this paper. 

By introducing magnetic charge into Maxwell's equations, dual-symmetry is obtained between the electric- and magnetic quantities of the theory. However, one cannot define a sensible dual-symmetry among the scalar and vector potential. In the work \cite{khademi2006non} (2006) another 4-vector is introduced into the dual-symmetric electromagnetic theory allowing a dual-symmetry to be defined among the potentials. In this description, the Dirac monopole has potentials without string singularities, since the magnetic field is no longer simply the curl of a vector potential.

For the interested reader, the dual-symmetric electrodynamics may be broken by introducing an axion-like pseudo-scalar field. Such a theory is considered in reference \cite{visinelli2011dual} (2011). 

Since the discovery of Dirac, several magnetic monopole like objects have been studied in the literature. In particular, magnetic monopoles appear in gauge theories, making them interesting for both particle physics and quantum field theory. It has been shown that all $SU(N)$ gauge theories contain magnetic monopoles as topological excitations of the spacetime vacuum labelled by the magnetic root lattice of $SU(N)$ \cite{englert1976quantization, goddard1978magnetic}. In the case of $N=2$, the Wu-Yang monopole is obtained.  A classical review of magnetic monopoles is found in Preskill \cite{preskill1984magnetic} (1984), and a more recent but less extensive review is found in Rajantie \cite{rajantie2012introduction} (2012). 

In 1974 't Hooft \cite{t1974magnetic} and Polyakov \cite{polyakov1974polyakov} showed that all Grand Unified Theories of particle physics has magnetic monopole solutions. Hence if nature is described by a Grand Unified Theory, magnetic monopoles must be a fundamental part of it. However, to this date (2019), no magnetic monopole particles have been observed. It is relevant to consider the monopole energy scale proposed by the Grand Unified Theories. This scale varies among the Grand Unified Theories, but generally depends on the unification energy of the theory. Under the assumption that no new physics happens between the experimental energy scale and the unification scale, one finds the monopole mass to be of order $10^{16}$ GeV \cite{kibble1976twb}. In this case, it is clear why no magnetic monopoles have been observed in the laboratory, as modern accelerator experiments at CERN reaches only a scale of $10^4$ GeV \cite{Restarting:1998739}. This made the way for cosmological magnetic monopole research \cite{kibble1976topology,zeldovich1978concentration,preskill1979cosmological}. Consider \cite{rajantie2012magnetic} for a recent (2012) brief review of magnetic monopoles in cosmology.

In the case of 't Hooft-Polyakov monopoles, the core radius is expected to be larger than the monopole Compton wavelength which implies semi classical behaviour. Hence studies of classical magnetic monopoles may contribute valuable information about the nature of such monopoles.

Although the 't Hooft-Polyakov monopoles have historically been drawing a lot of attention, a wealth of other magnetic monopoles have been studied. An important example in pure gauge theory is the Yang monopole \cite{yang1978generalization}, a five dimensional generalization of the Dirac monopole. The special feature of the Yang monopole is its non-vanishing second Chern number. In general, monopoles are characterised by a topological invariant related to their magnetic charge \cite{wu1975concept}. In five space dimensions, the topological invariant is the second Chern number. 

Recently monopoles have been simulated in the laboratory in ultracold atom experiments. The Yang monopole was realized in the laboratory by tuning the couplings between four hyperfine spin states in ultracold non-interacting bosonic atoms \cite{sugawa2016observation}. This was first done in 2016 by Sugawa \textit{et al.} at the National Institute of Standards and Technology (Gaithersburg, MD). Recently, an analysis of the interaction between multiple Yang monopoles was published \cite{yan2018yang} (2018). They find that different topological defects arise when multiple Yang monopoles are allowed to interact, and their results suggest that ultracold bosonic atoms may be used to create emergent topological defects. 

Another interesting class of monopoles are tensor monopoles. All monopoles mentioned above are examples of monopoles in vector gauge theories. Tensor gauge fields may be constructed as well and the analogue of Dirac's monopole in a tensor theory is constructed in references \cite{nepomechie1985magnetic,teitelboim1986monopoles,henneaux1986p}, the last in which electrodynamics is build on tensor gauge fields rather than the usual vector gauge fields.

It has been hypothesized that tensor monopoles may be realized in the laboratory as well. In a recent reference \cite{palumbo2018revealing} (2018) a three-level atomic system is proposed to realize a tensor monopole which maybe observed through quantum-metric measurements.

Magnetic monopoles appear and have applications in other aspects of physics than particle physics. A historically significant application of magnetic monopoles is its use in the study of quark confinement \cite{marciano1978quantum, mandelstam1983general}. In particular the Wu-Yang monopole has played a role \cite{ezawa1982abelian}. However, this monopole application has not been an active research field for the last couple decades, and the most recent reviews \cite{suzuki1995monopoles, yee1995monopoles} was published in 1995.

An interesting appearance of a magnetic monopole is found in reference \cite{moody1986realizations} (1986). Here it is shown that the effective Hamiltonian of diatomic nuclear motion is equivalent to that of a charged particle moving in a magnetic monopole field. In fact, this was made possible by earlier work of Berry \cite{berry1984quantal} (1984). Berry defined a gauge potential for adiabatic changes in quantum systems, and showed that monopoles appear in the parameter space of a two state system with a degeneracy point. More recently (2002), the Wu-Yang monopole has appeared in non-abelian plasma screening theory, the non-abelian generalization of Derby screening. The screening effects of a non-abelian plasma on a static chromoelectric field always gives a Wu-Yang monopole field at small distances \cite{jackiw1994hard, blaizot2002quark}. In reference \cite{stuller19802} the Wu-Yang monopole is used to calculate the Coulomb Green's function of $SU(2)$ Yang-Mills theory. The Green's function does not exist in general, as first pointed out by Gribov in \cite{gribov1978conferencia} (1978), hence the Wu-Yang monopole plays a special role in this context.

Recently a paper on $\mathcal{N}= 2$ supersymmetric pure gauge theory was published \cite{gillioz2017spontaneous} (2017). Here the Wu-Yang monopole appears as a solution with zero mass. The Wu-Yang like BPS monopole solves the same field equations \cite{prasad1975exact}.

\section{\label{sec:level1}The Wu-Yang monopole}

We start by introducing the Wu-Yang monopole, a classical monopole solution of the $SU(2)$ invariant Yang-Mills equations. These are the field equations obtained from a $SU(2)$ gauge invariant field theory. Consider Rubakov \cite{rubakov2009classical} for a reference on classical non-abelian gauge theory. The Wu-Yang solution was first mentioned by T.T. Wu and C.N. Yang in the paper \cite{wu1967some} in 1967. We consider a more general field configuration than that presented by Wu and Yang. The generalization is sometimes refereed to as the generalized Wu-Yang monopole, though we shall simply call it the Wu-Yang monopole.
For a general review of classical solutions of the $SU(2)$ Yang-Mills equations, consider the classical reference of Actor \cite{actor1979classical} (1979).

The Wu-Yang monopole is a mixing of $SU(2)$ gauge fields, $A_\mu^a$. The index $a=1,2,3$ denotes the three different gauge fields in a $SU(2)$ gauge field theory. One for each generator of the Lie Algebra. The fields are described by the Yang-Mills equations
\begin{align}
	\partial_\mu F_a^{\mu\nu}+\alpha\epsilon_{abc}A_\mu^b F^{\mu\nu}_c=0,
	\label{Yang-Mills}
\end{align}
where $F_a^{\mu\nu}=\partial^\mu A^\nu_a-\partial^\nu A^\mu_a+\alpha\epsilon_{abc}A^\mu_b A^\nu_c$ is the anti-symmetric field tensors. Here and throughout we shall use the
$(-,+,+,+)$ convention, and take Greek indices to run over 0,1,2,3 and Roman indices to run over 1,2,3.  Due to the nature of the Wu-Yang monopole we will not need the Lorentz sign convention as we go along. We will only need spacial indices. For the Levi-Civita symbol $\epsilon_{abc}$ we shall not distinguish between raised and lowered indices, e.g. $\epsilon_{abc}=\epsilon^a_{\,bc}$. The dimensionless constant $\alpha$ gives the coupling strength between the gauge fields. When $\alpha=0$ the theory describes three independent electromagnetic fields. For simplicity and without loss of generality we shall take $\alpha=1$. Because of the resulting non-linear term in equation \eqref{Yang-Mills} the three gauge fields $A^a_\mu$ interact mutually, i.e. they appear in each others field equations.

The Wu-Yang monopole is an ansatz of the field equations \eqref{Yang-Mills} of the form
\begin{align}
	&A^a_0=0,\quad A^a_i(\mathbf{x})=-\epsilon^a_{\, ij}x^j\frac{g(r)}{r^2}=\epsilon_{iaj}x^j\frac{g(r)}{r^2}, \\ 
	& a,i=1,2,3, \nonumber
	\label{Ansatz}
\end{align}
where $g(r)$ is a dimensionless spherical symmetric function. Then the gauge fields have units of inverse length dimensions as should be the case. We included the minus sign to follow the general sign convention found in the literature.

\begin{figure}
	\includegraphics[width=0.48\textwidth]{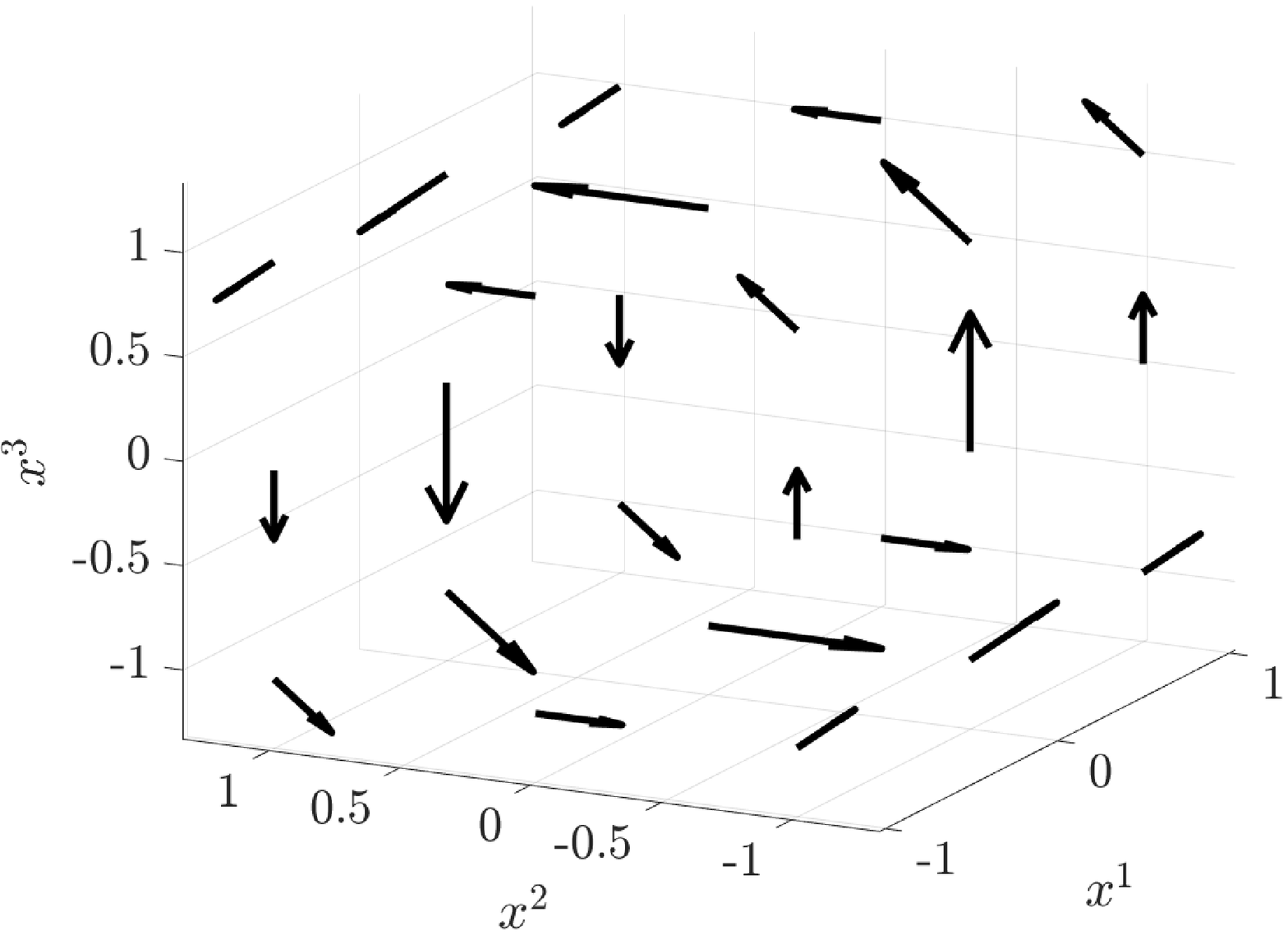}
	\caption{A vector plot of $\textbf{A}^1$ for the non-trivial vacuum $g=-2$. The plots of $\textbf{A}^2$ and $\textbf{A}^3$ looks the same, except the field $\textbf{A}^2$ rotates about $x^2$ and the field $\textbf{A}^3$ about $x^3$.\\ Here the axes are in units of some monopole length scale and the vectors are in inverse units of the same scale. At this point, we have no reason to choose any specific scale.}
	\label{fig1}
\end{figure}

It is clear why the configuration \eqref{Ansatz} is called a monopole. As it stands, it has a singularity at the origin and is scaled by a charge like function $g(r)$ called the monopole charge. The interesting feature of the Wu-Yang monopole is the distance dependence of the charge. This is different from the classical electric charge which is constant. Instead it reassembles the quantum mechanical distance dependence of charge which appears due to renormalization. The non-abelian feature of the Wu-Yang monopole is that, in oppose to a typical radial symmetric (e.g. electric) monopole field, it has a Levi-Civita mixing of space and gauge indices which breaks the overall radial symmetry. In figure \ref{fig1} it is clear that the vector field is not radial. Instead, it appears that $\textbf{A}^1=(A^1_1, A^1_2, A^1_3)$ rotates about the $x^1$-axis. In the same manner, $\textbf{A}^a$ rotates about the $x^a$-axis.

Contrary to what might be intuited from vector plots like Figure \ref{fig1}, the three gauge fields are not trivially related by rotations. Since $A^a_i(\mathbf{x})$ is a vector field it transforms as 
\begin{align}
	A^a_i(\mathbf{x})\rightarrow R_{ij} A^a_j(R\mathbf{x})
\end{align}
under a space rotation, $R$, which of course leaves the gauge index alone. It might be intuited that $\mathbf{A}^2(\mathbf{x})$ is obtained as a rotation of $\textbf{A}^1(\mathbf{x})$ about the $x^3$-axis (z-axis) by $\pi/2$ radians. However, doing the rotation one obtains
\begin{align}
&\textbf{A}^1(\textbf{x}) =(0, -x^3, x^2)\frac{g(r)}{r^2}\rightarrow \\ \nonumber& (x^3, 0, x^1)\frac{g(r)}{r^2}\neq\textbf{A}^2(\textbf{x})=(x^3, 0, -x^1)\frac{g(r)}{r^2},
\end{align}
which is not $\mathbf{A}^2(\mathbf{x})$. In order to obtain $\mathbf{A}^2(\mathbf{x})$, we need to also do a parity transformation of the $x^1$ axis: $x^1\rightarrow-x^1$. Thus the Gauge fields are not trivially related by rotations.

Notice that the Wu-Yang monopole naturally fulfils the Lorenz gauge condition as the interested reader may verify by the calculation found in Appendix \ref{Coulomb gauge}.

\subsection{\label{sec:action}The Wu-Yang action and field tensors}

In this subsection we state the field tensors and action implied by the Wu-Yang anzats \eqref{Ansatz}. In the next subsection we use the action to find the proper requirement on $g(r)$ in order that the $SU(2)$ Yang-Mills equations are fulfilled.

To find the Wu-Yang action we need the proper field tensors. These are rather complicated quantities calculated in Appendix \ref{Appendix1}:
\begin{align}
	&F^a_{ij}(\mathbf{x})=2\epsilon_{ija}\frac{g(r)}{r^2}+ \nonumber\\&[\epsilon_{jal}x^lx_i-\epsilon_{ial}x^lx_j]\left[\frac{g'(r)}{r^3}-2\frac{g(r)}{r^4}\right]+\epsilon_{ijl}x^lx^a\frac{g(r)^2}{r^4}.
\end{align}
Like the gauge fields, these field tensors are not radially symmetric. Interestingly, the non-radial contributions disappear in a lengthy calculation of the Lagrangian. This is anticipated, though, since the Lagrangian is expected to contain only information about the radial symmetric function $g(r)$, which is the only degree of freedom in the ansatz \eqref{Ansatz}. The calculation of the Lagrangian results in
\begin{align}
	L(r)&=-\frac{1}{4}F^a_{\mu\nu}F_a^{\mu\nu} \nonumber \\ &=-\left(\frac{g'(r)}{r}\right)^2-\frac{2g(r)^2+2g(r)^3+\frac{1}{2}g(r)^4}{r^4},
\end{align}
where we used the generic form of the $SU(2)$ Yang-Mills Lagrangian. Again, the calculation is found in Appendix \ref{Appendix2}.

Now we introduce a new function $f= g+1$. This deviates from Marinho \textit{et al.} \cite{marinho2009revisiting} (2010) who in our notation use $g=f+1$ to obtain the same results. This new function allows us to rewrite $L$ into the more compelling form
\begin{align}
	L(r)=-\left(\frac{f'(r)}{r}\right)^2-\frac{(f(r)^2-1)^2}{2r^4}.
	\label{Lagrangian}
\end{align}
This Lagrangian, called the Wu-Yang Lagrangian, is a considerable simplification of the generic Lagrangian, $L=-\frac{1}{4}F^a_{\mu\nu}F_a^{\mu\nu}$, which depends on all the gauge fields in a complicated manner. The problem of determining the free gauge fields has been reduced to that of determining a singe function of one variable. 

The action of the theory now takes the form
\begin{align}
S[f]&=-\int d^4x\ \left[\left(\frac{f'(r)}{r}\right)^2+\frac{(f(r)^2-1)^2}{2r^4}\right] \nonumber \\ &=-4\pi \int dt
 \int_0^\infty dr\ \left[f'(r)^2+\frac{(f(r)^2-1)^2}{2r^2}\right].
 \label{action}
\end{align}
This action is identical to that of an infinite long non-relativistic static string in a point-wise centrifugal potential about the origin. It has angular momentum squared $(f(r)^2-1)^2$, unit mass and string tension $8\pi$. Hence $f(r)$ may be thought of as a string displacement on a rigid sting. Each string constituent experiences a centrifugal force which depends on the displacement $f(r)$ from the string axis. Since the angular momentum depends on the displacement of the string rather than an angular velocity, the centrifugal potential cannot be thought of as coming from a spatial rotation. Notice that this interpretation has no relation to a Dirac string which is a feature of a diverging vector potential rather than a feature of a radially dependent monopole charge.

\subsection{The field equation\label{equation of motion}}

We now turn to the field equation for $f$. In accordance with the action principle, the action \eqref{action} is varied and required stationary. 

The variation with respect to $f$ gives
\begin{align}
\frac{\delta S[f]}{-4\pi \int dt}&=\int_0^\infty dr \left\{2f'\delta f'+2\frac{f^2-1}{2r^2}2f\delta f\right\} \nonumber \\&=
\int_0^\infty dr \left\{-2f''+2f\frac{f^2-1}{r^2}\right\}\delta f +\left.2f'\delta f\right|^\infty_0.
\label{vary}
\end{align}
To make the boundary term disappear we can take $f'$ or $\delta f$ to vanish at the boundaries. These are typical string boundary conditions, and are usually refereed to as a Neumann boundary condition and a Dirichlet boundary condition, respectively. A Neumann condition on both endpoints gives a free string and Dirichlet conditions gives a string with fixed endpoints.
In Subsection \ref{sec:level4} we will see that the condition $f'\rightarrow 0$ as $r\rightarrow \infty$ is natural as the energy contains an integral over $f'^2$. In other words, it is natural to have a free endpoint at infinity.

In any case, the boundary terms disappears and we obtain the condition
\begin{align}
	f''(r)=f(r)\frac{f(r)^2-1}{r^2}
	\label{Field equation}
\end{align} 
for a stationary action. From this equation we infer that each point along the string has curvature, $f''$, opposite directed to the displacement, $f$, as long as $|f|<1$. This is clear since such values of $f$ gives a negative fraction. Hence in such a domain, the structure of the string is oscillatory about the string axis defined by $f=0$. In fact, to first order in $f$ we obtain the equation \eqref{small_f} which resembles that of a simple harmonic oscillator with a radially dependent force constant equal to $r^{-2}$. We shall have more to say about this equation in Subsection \ref{sec:limitingsol}, for now simply notice that this implies that $f(r)=0$ can be thought of as a stable equilibrium for a sting constituent. Perturbing the string in either direction gives the nearby constituents an oscillating behaviour. Such oscillatory behaviour is illustrated in Figure \ref{fig:lin}.

When $f=0, \pm 1$ the curvature is zero, and when $|f|>1$ the curvature and position have identical signs resulting in the string curving away from the string axis. This tells that for $|f|>1$ the centrifugal force is strong enough to repel the string constituents from the axis. As we show in the proof of the Finite energy theorem in Appendix \ref{Appendix3}, the repulsion is strong enough to cause divergences in $f$. Figure \ref{fig:Taylor} illustrates the repulsion and beginning divergent behaviour of two solutions. One may think of $f(r)=\pm1$ as unstable equilibria of a string constituent, since perturbing it in either direction causes the nearby constituents to curve away from the string axis. In general, the size of curvature is strongly dependent on $r$ as it decreases like $r^{-2}$. Thus string constituents close to the origin tend to curve more than those further away. 

\subsection{\label{sec:level3}Properties of the field equation}

The field equation \eqref{Field equation} has some properties that will be important for what follows. One of these properties is parity symmetry. If $f$ is a solution then so is $-f$. This is seen as follows. Assume that $f$ is a solution, then
\begin{align}
(-f)''(r)=-\left[f(r)\frac{f(r)^2-1}{r^2}\right]=(-f)(r)\frac{(-f)(r)^2-1}{r^2},
\label{parity symmetry}
\end{align}
making $-f$ a solution.

Notice that parity symmetry is inherited from the Lagrangian \eqref{Lagrangian}. On physical grounds, one does indeed expect the string properties to be unchanged if each constituent is reflected in the string axis, as the angular momentum depends only on the distance from the string axis.

It is interesting to note that the Wu-Yang monopole \eqref{Ansatz} is not parity symmetric in $f$. This is clear from the constant solutions $f=0,\pm 1$ that we find in the next subsection. $f=1$ corresponds to vanishing gauge fields and $f=-1$ corresponds to a monopole field. This is a consequence of the equation relating $f$ to $g$, $f=g+1$, being dependent on parity.

The field equation also fulfils scale invariance, if $f$ is a solution then so is $h$ defined by $h(r)=f(\lambda r)$ for some dimensionless $\lambda> 0$. We choose $\lambda$ positive since $f(\lambda r)$ is not necessarily well-defined for negative $\lambda$, and for $\lambda=0$ the field equation breaks down. In fact, scale transformations are readily seen to define an equivalence relation among solutions, $f\sim h$. Hence the solutions on the Wu-Yang equation \eqref{Field equation} can be classified by a scale class and a parity.

To prove scale invariance we take $\rho =\lambda r$ and see that
\begin{align}
h''(r)&=\lambda^2 f''(\rho)=\lambda^2 f(\rho)\frac{f(\rho)^2-1}{\rho^2} \nonumber \\ & =h(r)\frac{h(r)^2-1}{r^2}.
\label{scale invariance}
\end{align}
Thus $h(r)$ fulfils the field equation.

Scale invariance allows the construction of arbitrary energy solutions from a particular non-zero finite energy solution. This is the scope of the Energy continuum theorem that we prove in Appendix \ref{Appendix4}.

\subsection{\label{solutions}Exact solutions of the field equation}

The field equation has no obvious closed solutions except the constant ones. If we assume that $f$ is a constant solution, then
\begin{align}
	0=f\frac{f^2-1}{r^2},
\end{align}
which is fulfilled when $f=0$ or $f=\pm1$. It is interesting to note the quantization in allowed constant solutions, they correspond to monopoles of constant charge $g=0,-1,-2$. Hence the three constant solutions give rise to three different configurations of the Wu-Yang monopole. The original monopole given by Wu and Yang corresponds to the generalized monopole of constant value $g=-1$. Notice the insignificance of the minus sign in the values of $g$. It may be absorbed by the Levi-Civita \eqref{Ansatz} by permuting a pair of indices.

As always, one can look for a power series solution of the equation \eqref{Field equation},
\begin{align}
&A(r)=\sum_{n=0}^{\infty}a_n r^n,\\
&B(r)=\sum_{n=0}^{\infty}b_n r^{-n}.
\end{align}
Here $A(r)$ represents those solutions which have a Tayler series expansion at $r=0$ and $B(r)$ the solutions which have an expansion at $r=\infty$. These expansions have been considered in the literature, consider Rosen (1972) \cite{rosen1972exact}, Actor (1979) \cite{actor1979classical} and for a discussion Marinho \textit{et al.} (2009) \cite{marinho2009revisiting}. We have supplemented the literature with a more detailed discussion of how to find the coefficients in Appendix \ref{Recursion formula}. As we show in this Appendix, the coefficients follow non-closed recursion relations with $a_0=0,\pm1$ and $b_0=0,\pm 1$. The solutions $a_0=0$ and $b_0=0$ both give the trivial constant solution mentioned above. The series for $A(r)$ contains only even powers of $r$ and has one free parameter $a_2=:a$. $a$ sets the energy scale of the solution as it has dimensions of energy squared (or inverse length squared). The series for $B(r)$ contains arbitrary powers of $r$ and has the free parameter $b_1=:b$. $b$ has units of energy and so sets the energy scale of the $B(r)$ solution.  The first 10 coefficients of $A(r)$ and $B(r)$ are given in Marinho \textit{et al.} \cite{marinho2009revisiting}. We do not agree with their assertion of signs(neither do Rosen \cite{rosen1972exact} or Actor \cite{actor1979classical}). For $a_0=1$ we state here the more general form of $A(r)$:
\begin{align}
&A(r)=1+ar^2+ \frac{3}{10}a^2r^4+\frac{1}{10}a^3r^6+ \frac{59}{1800} a^4r^8\nonumber \\&+ \frac{71}{6600}a^5r^{10} + \frac{15143}{4290000} a^6r^{12} + \frac{20327}{17550000} a^7r^{14} \nonumber \\&+ \frac{1995599}{5250960000} a^8r^{16}
+\frac{ 311031533 }{2494206000000}a^9r^{18}.
\label{expansionA}
\end{align}
A natural question arises, how is $a_0$ put back into equation \eqref{expansionA}. Using parity symmetry \eqref{parity symmetry}, we have that $-A(r)$ is a solution with all signs reversed. There is still a freedom in signs, though, for the uneven powers in $a$ one can do the transformation $a\rightarrow -a$. Thus it is clear that only terms with even powers of $a$ needs a factor of $a_0$ to change signs under a parity transformation. This argument is supported by our explicit calculations in Appendix \ref{Recursion formula}. Notice that the expansion \eqref{expansionA} contains the constant solutions $A(r)=\pm 1$ by choosing $a=0$. In Figure \ref{fig:A(r)andf(r)} the graph of $A(r)$ is found together with the graph of the corresponding numerical solution. A near perfect correspondence is found when $r\leq1.4a$. The numerical solution if plotted again in Figure \ref{fig2a} where more details are included.

\begin{figure}
	\centering		
	\includegraphics[width=0.50\textwidth]{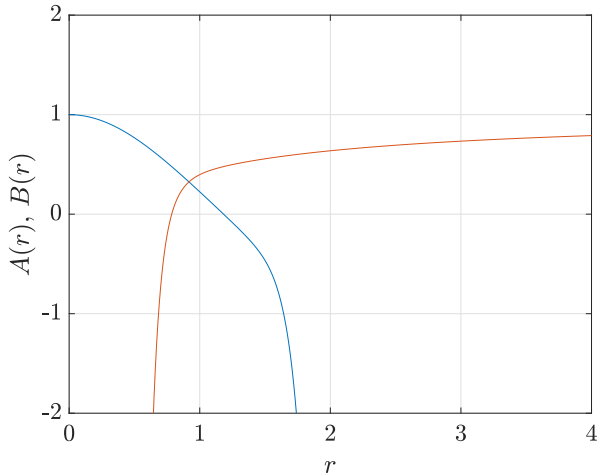}
	\caption{A plot of the expansions $A(r)$ from equation \eqref{expansionA} (blue graph to the left) and $B(r)$ from equation \eqref{expansionB} (orange graph to the right) with $b=-1$, $a=-1$ and $b_0=1$. The radial axis is given in units of $(-a)^{-1/2}=-b$. From the figure it is clear where the domains of convergence overlap. It appears the graphs cannot be joined in a differentiable manner. Notice the divergent behaviours at the endpoints of the domains of convergence given in equations \eqref{convergenceA} and \eqref{convergenceB}. \label{fig:Taylor}}
	\end{figure}

\begin{figure}
	\centering		
	\includegraphics[width=0.50\textwidth]{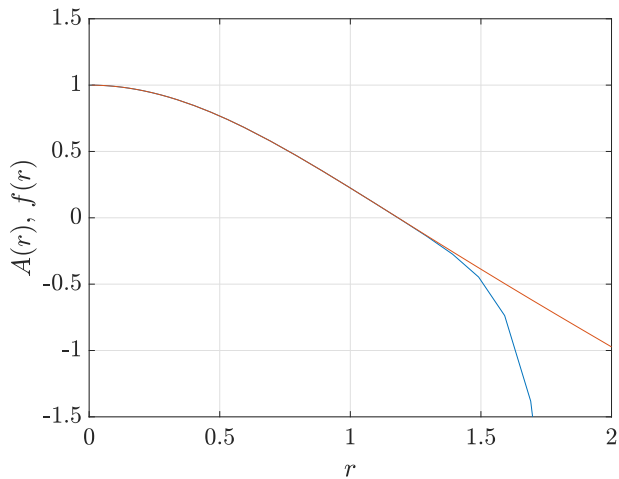}
	\caption{Here the Taylor expansion $A(r)$ from equation \eqref{expansionA} (blue lower graph) with $a=-1$ is compared with the corresponding numerical solution (orange upper graph). The radial axis is in length units $(-a)^{-1/2}$. Notice the near perfect agreement between the graphs when $r\leq1.4(-a)^{-1/2}$. \label{fig:A(r)andf(r)}}
\end{figure}

The corresponding expansion for $B(r)$ is more complicated as the uneven coefficients does not vanish in general. By the same considerations as above we have
\begin{align}
	&B(r)=b_0 +b\frac{1}{r}+ b_0b^2\frac{3}{4}\frac{1}{r^2} +b^3\frac{11}{20}\frac{1}{r^3} + b_0b^4\frac{193}{480}\frac{1}{r^4} \nonumber \\ &+b^5\frac{47}{160} \frac{1}{r^5}
	+b_0b^6 \frac{3433}{16000}\frac{1}{r^6} + b^7\frac{67699}{432000}\frac{1}{r^7} \nonumber \\ & + b_0b^8\frac{1318507}{11520000}\frac{1}{r^8}+b^9\frac{2118509}{25344000}\frac{1}{r^9}.
	\label{expansionB}
\end{align}
Notice that for both expansions, a scale transformation \eqref{scale invariance} corresponds to redefining the free parameters $a$ and $b$. The graphs of $A(r)$ and $B(r)$ are shown in Figure \ref{fig:Taylor}.

From the higher order terms of the above partial sums of $A(r)$ and $B(r)$, we can estimate the radii of convergence using the ratio test. For $A(r)$ we have convergence when
\begin{align}
\left|\frac{ 311031533 }{2494206000000}a_0a^9r^{18}/\frac{1995599}{5250960000} a^8r^{16}\right|<1,
\end{align}
which, using $|a_0|=1$, reduces to
\begin{align}
	\sqrt{a}r<1.746.
	\label{convergenceA}
\end{align}
For $B(r)$ we find convergence for
\begin{align}
	\frac{r}{b}>0.730.
	\label{convergenceB}
\end{align}
According to Marinho \textit{et al.} \cite{marinho2009revisiting}, these estimates of convergence changes little when using higher order coefficients. We find that neither of the power series converge for all $r>0$, but each may come arbitrarily close by choosing $\sqrt{a}$ very small or $b$ very large in some particular length units. 

We now ask if the solutions may be glued together at some length $0.73b<R<\frac{1.75}{\sqrt{a}}$ creating a smooth solution of the field equation \eqref{Field equation},
\begin{align}
f(r)=
\begin{cases}A(r)=\sum_{n=0}^{\infty}a^{n} r^{2n}, & 0\leq r \leq R, \\
B(r)=\sum_{n=0}^{\infty}b^n r^{-n}, & R\leq r.
\end{cases}
\label{combined power series}
\end{align}
Here we have chosen $a_0=1=b_0$ for concreteness. In this case $a^{-1/2}$ plays the role of the small scale length of the problem, the radius where the zeroth order term of $A(r)$ starts to dominate. Likewise, $b$ plays the role of a large scale length determining when the zeroth order term of $B(r)$ starts to dominate. An example of a combined power series solution \eqref{combined power series} is seen in Figure \ref{fig:Taylor} when the diverging parts are ignored.

Equation \eqref{combined power series} cannot be constructed smoothly as we now show. By assuming that $A(R)=B(R)$ we obtain the equation
\begin{align}
0=\sum_{n=0}^\infty(a^nR^{2n}-b^nR^{-n})=\sum_{n=0}^{\infty}b^nR^{-n}\left(\frac{a^n}{b^n}R^{3n}-1\right),
\end{align}
which is fulfilled if $b=aR^3$. Considering the analogous equation for $A'(R)=B'(R)$ we arrive at
\begin{align}
	0&=\sum_{n=1}^{\infty}(2na^nR^{2n-1}+nb^nR^{-(n+1)})\nonumber \\&=\sum_{n=1}^{\infty}nb^nR^{-(n+1)}\left(2\frac{a^n}{b^n}R^{3n}+1\right),
\end{align}
which is fulfilled if $b=-2aR^3$. It is clear that both conditions cannot be fulfilled simultaneously, thus we see that $A(r)$ and $B(r)$ cannot be joined smoothly. In Section \ref{sec:level5} this becomes clear from energy considerations. It is not without physical motivation, however, to go on an study solutions which have a discontinuity in the first derivative. It turns out (see Section \ref{sec:level5}) that such solutions are the only non-trivial solutions which have the possibility of finite energy. We study one such solution in detail in Subsection \ref{sec:level8} and Marinho \textit{et al.} \cite{marinho2009revisiting} considers four such solutions.

Allowing a discontinuity in the derivative of size $f'(R_+)-f'(R_-)=\zeta$ at some radius $R$ corresponds to adding a delta function to the field equation \eqref{Field equation},
\begin{align}
	f''(r)=f(r)\frac{f(r)^2-1}{r^2}+\zeta \delta(r-R),
\label{modified field equation}
\end{align}
since for $\epsilon>0$ we have
\begin{align}
	&f'(R_+)-f'(R_-) \nonumber \\
	&=\lim_{\epsilon\rightarrow 0}\int_{R-\epsilon}^{R+\epsilon}dr\left[f(r)\frac{f(r)^2-1}{r^2}+\zeta \delta(r-R)\right]=\zeta.
	\label{Zeta}
\end{align}
Here the first term on the second line vanish due to continuity. 

Notice that the introduction of the delta function and the scales $R$ and $\zeta$, implies that the scale transformed function $f(\lambda r)$ is a solution with scales $R/\lambda$ and $\lambda\zeta$:
\begin{align}
	\frac{d^2}{dr^2}f(\lambda r)&=\lambda^2\left[f(\lambda r)\frac{f(\lambda r)^2-1}{(\lambda r)^2}+\zeta \delta(\lambda r-R)\right] \nonumber \\
	&=f(\lambda r)\frac{f(\lambda r)^2-1}{r^2}+\lambda\zeta \delta(r-R/\lambda).
\end{align}
As such, the introduction of a discontinuity in the derivative does not break scale invariance, except if the scales $R$ or $\zeta$ are taken to be universal. See Marinho \textit{et al.} \cite{marinho2009revisiting} for a discussion of \eqref{modified field equation} in the context of QCD. They suggest that the delta function may be thought of as localized static quarks at radius $R$ with $\zeta$ as the square of the quark wave function. In QCD the relevant (universal) length scale for breaking scale symmetry is of order $R\sim 1\, \mathrm{fm}$.

Next, we considered the novel approach of using Pad{\'e} approximants in the field equation \eqref{Field equation}. A Pad{\'e} approximant is an ansatz of the form
\begin{align}
	f(r)=\frac{\sum_{m=0}^{M}a_mr^m}{1+\sum_{n=1}^{N}b_nr^n}.
\end{align}
In Appendix \ref{Pade} we find the restrictions imposed by the field equation on the first four coefficients. We find that like for the power series expansions, $a_0=0,\pm1$. For the next three coefficients we find that $a_n=a_0b_n$. Assuming this trend continues, we have that Pad{\'e} approximants give no new information, they simply reproduce the constant solutions $f=0,\pm1$.

\subsection{Limiting solutions of the field equation \label{sec:limitingsol}}

Another class of analytical solutions are those found in the extreme cases $|f(r)|\ll 1$ and $|f(r)\gg1$. Unfortunately, as we will see, such solutions does not exist globally.

The linearised field equation is on the form
\begin{align}
	f''(r)=-\frac{1}{r^2}f(r),
	\label{small_f}
\end{align}
and is fulfilled by a solution when $|f(r)|\ll 1$. It is curious to notice that the linearised field equation is equivalent to the time-independent Schrödinger equation with vanishing energy in a $-1/{r^2}$ potential. The solutions have been studied in detail by Landau and Lifshitz in their work on quantum mechanics \cite{landau2013quantum}. This equation has an intriguing application in the Efimov effect \cite{efimov1970energy}. The Efimov effect happens in a system of three identical bosons, when a two body subsystem is at the dissociation energy. Under these conditions, surprisingly, there is an infinite number of bound states. For a general review of the three body problem with short-range interactions, consider Nielsen \textit{et. al} \cite{nielsen2001three}.
In the context of the Wu-Yang monopole consider Appendix G of Actor \cite{actor1979classical}.

It is readily checked that \eqref{small_f} has solutions on the form
\begin{align}
f_l(r)=\sqrt{r/r_0}\left[A\cos\left(\frac{\sqrt{3}}{2}\ln r/r_0\right)+B\sin\left(\frac{\sqrt{3}}{2}\ln r/r_0\right)\right].
\label{Linearised}
\end{align}
Here $r_0$ is the length scale where $|f(r)|$ becomes comparable to one and the linear approximation breaks down. 

\begin{figure}
	\centering		
	\includegraphics[width=0.50\textwidth]{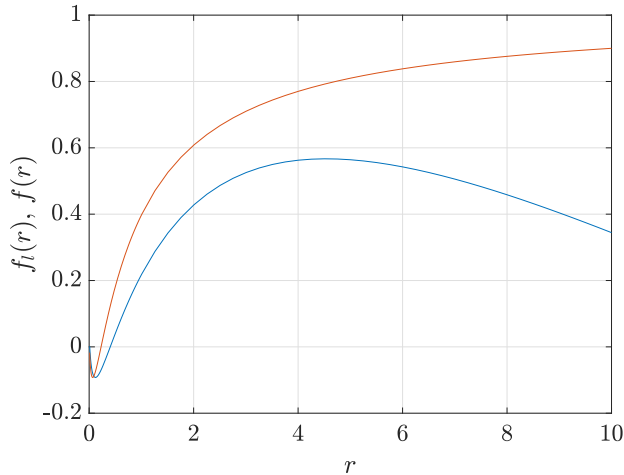}
	\caption{Here the linearised solution $f_l(r)$ from equation \eqref{Linearised} (blue lower graph) with $A=B=1$ and $r_0=4.6a$ is compared to the corresponding numerical solution (orange upper graph).  The radial axis is in some length units $a$. The agreement between the graphs is good for small $r$. However, already at the first minimum the graphs disagree. $f_l$ has minimum at a larger radius than $f$ with a difference of $0.1a$.}\label{fig:lin}
\end{figure}

The solution is plotted as the blue graph in Figure \ref{fig:lin} for $A=B=1$ and $r_0=4.6$. As expected by expression \eqref{Linearised}, the function oscillates while getting an increasingly lager amplitude. The frequency is radially dependent through the logarithm of $r$, which is also seen in the figure. It appears that already at the first minimum, $f_l(r)=-0.1$, the assumption $|f_l(r)|\ll 1$ is broken. This is definitely the case at the maximum which appears around $r=4a$ with the value $f_l(r)=0.55$. To make a comparison, we found the numerical solution with the corresponding behaviour at small $r$. This is seen in Figure \ref{fig:lin} as the orange graph, where the first minimum resembles that of the linearised solution. The maximum is nowhere to be found, it is replaced by a monotonic growth. Hence it is clear that the solution of the linearised field equation is a good approximation of the numerical solution only for radii much smaller than the length scale $a$. However, It might be that $f_l$ approximate solutions of the field equation at each minima.

In the other limit when $|f(r)|\gg1$ the $f$ cubed term dominates and the field equation takes the form
\begin{align}
	f''(r)=\frac{f(r)^3}{r^2}.
	\label{Cubed}
\end{align}
This equation has a solution in the same $|f(r)|\gg1$ limit. Consider the function
\begin{align}
	f_c (r)=\frac{\sqrt{2}r_0}{r_0-r}
	\label{cubed solution}
\end{align}
for some diverging length scale $r_0$. In the limit $|f_c(r)|\gg1$ it is true that $|r_0-r|/r_0\ll1$. Hence to lowest order in $(r_0-r)/r_0$ we have
\begin{align}
	\frac{f_c(r)^3}{r^2}=\frac{2\sqrt{2}r_0^3}{[r_0+(r-r_0)]^2(r-r_0)^3}=\frac{2\sqrt{2}r_0}{(r-r_0)^3}=f_c''(r),
\end{align}
and so $f_c$ is a solution of the cubed field equation \eqref{Cubed} when $|f_c(r)|$ is large. This implies that whenever a solution of the field equation \eqref{Field equation} grows large, the Wu-Yang monopole has a divergence. As we will see, the energy of the Wu-Yang monopole depends on the space integral of $f^4/r^2$ and the behaviour of $f_c$ gives an infinite energy. This is consistent with the Finite energy theorem (see Appendix \ref{Appendix3}) which states that whenever there is a region where $|f(r)|>1$, the energy diverges. The main argument in the proof is that when $|f(r)|>1$ for some radius $r$, $|f|$ grows without bound in either the increasing or decreasing radial direction. The behaviour of this diverging growth is captured by \eqref{cubed solution}.

\subsection{\label{sec:level4}The field energy}

The Wu-Yang gauge fields are time-independent, hence the energy density, $\mathcal{E}=\frac{\partial L}{\partial \dot{A}_\mu}\dot{A}_\mu-L$, equals the negative Lagrangian. In spherical coordinates this gives the energy formula
\begin{align}
	E=4\pi\int_0^\infty dr\ \left[f'^2+\frac{(f^2-1)^2}{2r^2}\right],
	\label{Energy}
\end{align}
where we used the Lagrangian from equation \eqref{Lagrangian}. The energy is positive definite since all terms are squared. This gives a lower bound on the energy as required of a stable theory of physics. We notice that the energy integral is indeed that of a static infinite string in a centrifugal potential.

The energy is minimized when equation \eqref{Energy} vanishes. This is fulfilled when
\begin{align}
\left[f'^2+\frac{(f^2-1)^2}{2r^2}\right]=0.
\end{align}
Both terms are non negative so the equation holds true if and only if both terms are zero. Therefore $f^2-1=0$ making
\begin{align}
f=\pm 1
\end{align}
the only vacuum configurations. These are straight strings, each lying on a minimum line of the centrifugal potential surface. These lines corresponds to vanishing centrifugal potential for all string constituents. As discussed earlier in relation to the field equation, each point on these lines corresponds to a locally unstable equilibrium. However, this shows that when all parts of a string are located at an unstable equilibrium, the string as a whole becomes stable.

We may rewrite the energy integral if we consider a free string, that is a string with Neumann boundary conditions, $f'\rightarrow0$ as $r\rightarrow 0,\infty$. We rewrite the kinetic term of the energy by partial integration and use the field equation \eqref{Field equation} which holds for physical strings:
\begin{align}
	E&=4\pi\int_0^\infty dr\ \left[-ff''+\frac{(f^2-1)^2}{2r^2}\right] \nonumber \\
	&=4\pi\int_0^\infty dr\ \left[-f^2\frac{(f^2-1)}{r^2}+\frac{(f^2-1)^2}{2r^2}\right] \nonumber \\
	&=-2\pi\int_0^\infty dr\ \left(f^2+1\right)\frac{f^2-1}{r^2}.
	\label{Energy2}
	\end{align}

This gives an alternative expression of the energy density of a free string. The result has surprising implications. Since the energy is positive definite for all $f$, equation \eqref{Energy2} implies that the relationship between the domain, $D_+=D_{oscillating}$, where the new energy density is positive, i.e. $f^2<1$, and the domain, $D_-=D_{diverging}$, where the new energy density is negative, i.e. $f^2>1$, is such that the energy integral turns out positive. If the energy is zero, then constantly $f^2=1$ in agreement with our previous result. This relation between domains is interpreted as follows. For every part of the string with a diverging appearance, there must be a corresponding part with an oscillating appearance (see Subsection \ref{equation of motion} for terminology). However, there may be oscillating parts without corresponding diverging parts. Mathematically we have the inequality:
\begin{align}
	\int_{D_+} dr\ \left(f^2+1\right)\frac{f^2-1}{r^2}\geq \int_{D_-} dr\ \left(f^2+1\right)\frac{1-f^2}{r^2}.
	\label{Energy inequality}
\end{align}

This result is quite intriguing in the light of the Finite energy theorem of Appendix \ref{Appendix3}, which states that for a smooth string, any constituent with $|f(r)|>1$ gives rise to an energy divergence in a subset of $D_-$. This implies that when the right hand side of the above inequality is non-zero, then it is infinite. Due to the inequality, this gives rise to the surprising conclusion, that if a string has diverging parts then its oscillating parts also diverge the energy. This result makes it natural to restrict attention to solutions without repulsive parts, making $D_-$ the empty set and the integral on the right vanish. However, we find in Section \ref{sec:level5} that no matter what is done, any smooth solution has a divergent energy. 

\subsection{Vacuum configurations}

It is interesting to consider the actual gauge fields, $A_i^a$, corresponding to the vacuum solutions $f=\pm1$. From \eqref{Ansatz} and $g=f-1$ we find that when $f=1$ the gauge fields has monopole charge $g=0$, giving the trivial vacuum:
\begin{align}
	A_i^a=0.
	\label{trivial_vacuum}
\end{align}
However when $f=-1$ we obtain the monopole of constant charge $-2$:
\begin{align}
	A_i^a=\epsilon^a_{\;ij}x^j\frac{2}{r^{2}},
	\label{non-trivial_vacuum}
\end{align}
seen in Figure \ref{fig1}. Interestingly, this shows the existence of a vacuum configuration with non-vanishing gauge fields. 

The two vacuum configurations of the Wu-Yang monopole are quite different, the one being trivial and the other having a singularity. As mentioned earlier, this is an example of the curiosity that the parity symmetry of the string is not carried by the Wu-Yang monopole \eqref{Ansatz}.

We now ask whether it is possible to make a physical transition from one vacuum to the other. This cannot happen in a continuous way using only a finite amount of energy. To see this, consider the continuous deformation of the vacuum string $f=1$ to a new string with $f(0)=1-\delta$. This new string has infinite energy since the centrifugal contribution of \eqref{Energy} gives a divergence in the limit $r\rightarrow 0$. Hence we find the vacuum configurations are separated by an infinite energy barrier and no continuous transitions are allowed between them. The only possible transitions between the two vacuums are discrete ones. 

Notice the curious fact that the last constant solution, $f=0$, carries infinite energy,
\begin{align}
E=2\pi\int_0^\infty dr\frac{1}{r^2}=\infty.
\end{align}
This result is not anticipated when considering the field equation which has $f(r)=0$ as a stable equilibrium for a string constituent at $r$. However, the energy formula has a non-vanishing $r^{-2}$ contribution from the centrifugal term which diverges the energy and thus makes the string globally unstable.

\subsection{The Wu-Yang monopole as a collection of magnetic dipoles}

We conclude this section by considering the non-abelian generalizations of electric- and magnetic fields in the case of the Wu-Yang monopole. We anticipate the Wu-Yang monopole to have non-zero magnetic fields since each of the gauge fields exhibits a magnetic like rotation, as illustrated in Figure \ref{fig1}. Recall that usually the magnetic field is related to rotations in the vector field by $\textbf{B}=\mathrm{curl}\textbf{A}$. Furthermore, $g$ is anticipated to play the role of a radial dependent magnetic charge due to its role as a monopole charge in the structure of the Wu-Yang monopole \eqref{Ansatz}. However, it turns out that the magnetic structure of the solution is not monopole like, but rather dipole like depending on both $g$ and the derivative $g'$.

Inspired by electromagnetism, we introduce the non-abelian electric- and magnetic fields as follows:
\begin{equation}
E^a_i=F^a_{i0},\qquad B^a_i=\frac{1}{2} \epsilon_{ijk}F^a_{jk}.
\end{equation}
We quickly realize that these fields are not gauge invariant in a non-abelian theory where the field tensor transforms under gauge transformations. Hence they are not observables. Rather, they are theoretical constructions which hints at the properties of a non-abelian field.

We calculate the chromomagnetic fields in Appendix \ref{Magnetic field}. It is found that
\begin{align}
	B^a_i&
	=2\delta_{i}^a\frac{g(r)}{r^2}+[\delta_{i}^ar^2-x_ix^a]\left[\frac{g'(r)}{r^3}-2\frac{g(r)}{r^4}\right].
	\nonumber \\&\quad+x_ix^a\frac{g(r)^2}{r^4}.
	\label{Magnetic}
\end{align}
The electric fields vanish due to time-independence and the $A^a_0=0$ assumption of the ansatz, thus $E^a_i=F^a_{i 0}=0$.

\begin{figure}
	\centering		
	\includegraphics[width=0.50\textwidth]{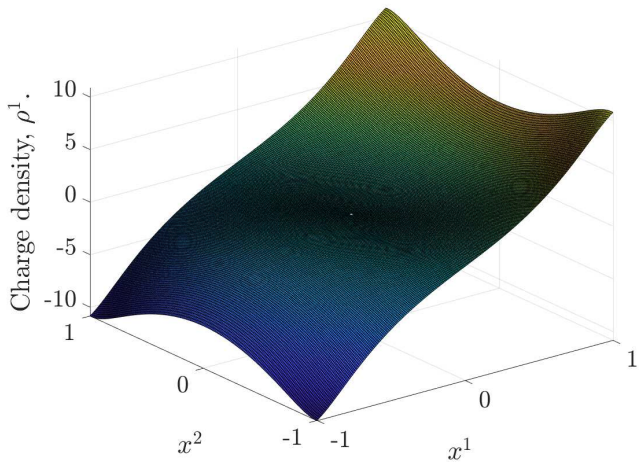}
	\caption{A surface plot of the first component, $\rho^1$, of the charge density \eqref{chargedensity} in two space dimensions. The power series expansion \eqref{expansionA} with $a=-1$ is used as monopole charge. The $x^1$ and $x^2$ axes are given in units of $(-a)^{-1/2}$, and the charge density in units of $(-a)^{3/2}$.}\label{fig:surface}
\end{figure}

\begin{figure}
	\centering		
	\includegraphics[width=0.50\textwidth]{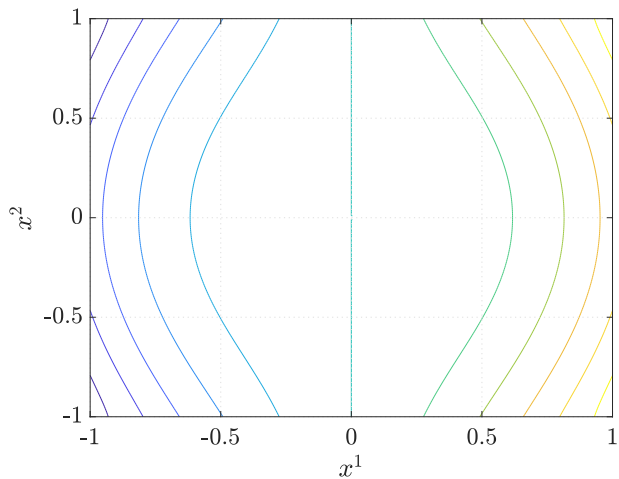}
	\caption{A two dimensional contour plot of the charge density $\rho^1$ from equation \eqref{chargedensity}. We used the power series expansion \eqref{expansionA} with $a=-1$ as the monopole charge. The axes are given in units of $(-a)^{-1/2}$.}\label{fig:contour}
\end{figure}

The form of $B^a_i$ shows that in a complicated manner, $g$ plays the role of a radial dependent magnetic charge, in the sense that it determine the strength of the magnetic fields. This was to be expected, after all $g$ is the only degree of freedom in the Wu-Yang monopole. Interestingly, the derivative of the charge plays a role too. In order to be concrete, inspired by electromagnetism, we calculate the divergence of the magnetic fields $\textbf{B}^a$ to find a result proportional to the associated charge densities. This calculation is done in Appendix \ref{Magnetic field}, where the complicated expression \eqref{Magnetic} is shown to have a surprisingly simple divergence. The magnetic charge densities are
\begin{align}
	\rho^a(x)=\partial_i B^a_i(x)=2x^a\frac{g(r)g'(r)}{r^3}.
	\label{chargedensity}
\end{align}
First of all these are symmetric in the gauge index $a$, related by rotations of $\pi/2$ radians. They are the charge densities of three dipoles aligned along one axis each. In this sense, the Wu-Yang monopole is not a magnetic monopole, but rather consists of three magnetic dipoles with no net charge, as is a consequence of the symmetry $\rho(x)\rightarrow -\rho(x)$ when $x\rightarrow-x$. As expected from the appearance of the magnetic field, $g'(r)$ plays a role along with $g(r)$. What was not expected is that $g'(r)$ plays a role on equal footing with $g(r)$. In Figure \ref{fig:surface} and \ref{fig:contour} the first component of the charge density is plotted using the power series solution \eqref{expansionA} with $a=-1$. The plot uses only two space dimensions, but the permutation symmetry in $x^2$ and $x^3$ gives the behaviour in the $x^3$ direction.
It is intriguing, that a source-free self-interacting theory like the present one can be understood as creating its own charges interacting with its own gauge fields via analogues of Maxwell's equations. This is a clear and concrete representation of the self-interaction of a non-abelian gauge theory. However, as was argued, no net charge is created within the set-up.

To conclude this section let us consider the magnetic fields and charge densities of the vacuum configurations. They are expected to vanish since magnetic fields and charges are sources of energy. Indeed this is what we find. Both solutions $g=0,-2$ have a vanishing derivative implying that $\rho^a=0$. For the trivial vacuum we have $g=0$ and $g'=0$ implying that $B^a=0$. For the non-trivial vacuum, $g=-2$ and thus the derivative term in $B^a$ vanishes. A little calculation shows that the resulting terms cancel in pairs.

\section{\label{sec:level5}Finite energy solutions and non-existence}

In this section we go into energy considerations of the Wu-Yang monopole. We argue that no smooth finite energy solutions exist that are not trivial. Here the word trivial is synonymous with vanishing gauge fields. We argue that no smooth non-trivial solutions exist even if we allow them to be singular. For this, we take our starting point in a non-existence theorem given in the paper ''There are no classical glueballs'' (1977) by Coleman \cite{coleman1977there}. The non-existence theorem states that no smooth non-trivial finite-energy non-singular solutions of classical Yang-Mills theory in four dimensional Minkowski space exists except if they radiate energy toward spatial infinity. To be concrete, Coleman \cite{coleman1977there} does not mention smoothness explicitly, going through the arguments, one finds that two times continuous differentiable gauge fields are assumed. We have not been able to find even one time differentiable solutions of finite energy. An example is the gluing together of the power series solutions \eqref{combined power series} in Subsection \ref{solutions}. This could not be done in a differentiable manner, but the continuous solution turns out to have finite energy. In Subsection \ref{sec:level8} we give a numerical solution with discontinuous derivative, this is a solution of the modified field equation \eqref{modified field equation}. Several piecewise smooth solutions are given in Marinho \textit{et al.} \cite{marinho2009revisiting}.

\subsection{\label{sec:level6}A non-existence theorem}

Here we state the non-existence theorem, deduce the corollary it imposes on the Wu-Yang monopole and discuss its implications. 

The non-existence theorem in \cite{coleman1977there} states that no smooth non-trivial finite-energy non-singular solutions of classical Yang-Mills theory in four dimensional Minkowski space exists except those that radiate energy toward spatial infinity. For a static solution, like the Wu-Yang monopole, it is not possible to radiate. Hence we obtain the corollary that no smooth non-trivial non-singular static solutions of finite energy exists.

Thus the theorem implies that any smooth non-trivial finite energy Wu-Yang solution has to be singular. In other words, we cannot find any non-trivial solutions of the field equation \eqref{Field equation} which gives both a non-divergent gauge field configuration and a non-divergent energy. This is a fundamental difficulty of the Wu-Yang monopole since true divergent behaviour is not observed in nature. It so turns out, that one cannot find even singular smooth non-trivial finite energy solutions as we show in the next subsection. Thus as it will turn out in Subsection \ref{sec:level8}, for the Wu-Yang monopole the essential assumption of the non-existence theorem is smoothness. If we let go of this assumption, we are able to find several different classes of finite energy solutions.

The non-existence theorem is consistent with the constant solutions that was obtained in Subsection \ref{solutions}. The vacuums ($f=\pm1$) have finite energy which is consistent since the monopole \eqref{Ansatz} with $f=1$ is trivial and with $f=-1$ is singular. The constant solution $f=0$ has both infinite energy and a singular monopole.

\subsection{\label{sec:level7}Conditions for non-divergent energy}

In this subsection we consider what conditions are needed in order not to diverge the energy. We prove in Appendix \ref{Appendix3} the Finite energy theorem stating that any smooth finite energy solution, $f$, must have $|f(r)|\leq1$ for all $r$. The key point in the proof is that anywhere $|f(r)|>1$, the function grows unbound and reach the asymptotic solution \eqref{cubed solution} which diverge the energy. The Finite energy theorem is argued in Marinho \textit{et al.} \cite{marinho2009revisiting}. If smoothness is not required of the solution, the Finite energy theorem is invalid, but still serves as a statement of what natural behaviour a finite energy solution have. For an intriguing application of the Finite energy theorem consider the discussion at the end of Subsection \ref{sec:level4}. 

When $|f(r)|\leq1$ for all radii, it is clear from the energy formula \eqref{Energy} that divergence can happen only in the limits $r\rightarrow 0$ and $r\rightarrow \infty$. We find that a solution with
\begin{align}
	f(r)\rightarrow [1- (\sqrt{a} r)^n]\quad \mathrm{as}\quad r\rightarrow 0,
	\label{asymptoticrsmall}
\end{align}
must have $n>1/2$, and a solution with
\begin{align}
f(r)\rightarrow \left[1- \left(\frac{b}{r}\right)^n\right]\quad \mathrm{as}\quad r\rightarrow \infty,
\label{asymptoticrlarge}
\end{align}
must have $n>0$ in order to give a finite energy contribution in the limits. However, as we show below, for the behaviour in \eqref{asymptoticrsmall} the smallest $n$ which solves the field equation \eqref{Field equation} is $n=2$, and for the behaviour in \eqref{asymptoticrlarge} the smallest $n$ solving the field equation is $n=1$. These are the lowest order (non-trivial) terms we already know from the power series expansion \eqref{combined power series} that was considered in Subsection \ref{solutions}. As such, it is clear that the non-smooth combined solution \eqref{combined power series} has finite energy. Remember that due to parity symmetry \eqref{parity symmetry} we could have considered $-f(r)$ as well in \eqref{asymptoticrsmall} and \eqref{asymptoticrlarge}.

We now deduce the condition $n>1/2$ on the asymptotic behaviour in equation \eqref{asymptoticrsmall}. Considering first the centrifugal term of the energy integral \eqref{Energy}, it is clear that we obtain a divergence if not $f(r)\rightarrow \pm 1$ fast enough as $r\rightarrow 0$. Plugging \eqref{asymptoticrsmall} into the energy integral \eqref{Energy} we find
\begin{align}
	E^\delta[f]&=4\pi \int_{0}^{\delta}dr\ \left[a^{n} r^{2(n-1)}+\frac{\left((1-a^{n/2} r^n)^2-1\right)^2}{2r^2}\right] \nonumber \\
	&=4\pi \int_{0}^{\delta}dr\ \left[a^{n} r^{2(n-1)}+\frac{\left(a^{n} r^{2n-1}-a^{n/2} r^{n-1}\right)^2}{2}\right],
\end{align}
where $\delta$ is some small length in the sense that $\delta\sqrt{a}\ll1$, making the leading order assumption on $f$ in equation \eqref{asymptoticrsmall} a good approximation. The first term in the integrand requires that $n>\frac{1}{2}$. Such $n$ gives the first term inside the second term bracket a positive power, and the second term a power strictly larger than $-\frac{1}{2}$. Hence the second term converges when $n>\frac{1}{2}$ making it a proper requirement on $n$. 

We know from the power series expansion at $r=0$ \eqref{expansionA} that the lowest order integer exponent compatible with the field equation \eqref{Field equation} has $n=2$. This gives an asymptotic solution with finite $E^\delta$. What we do not know is if the field equation permits solutions of non-integer $n$ which does not diverge $E^\delta$. This we consider know.

Plugging into the field equation \eqref{Field equation} we find in the limit $r\rightarrow 0$ the left hand side
\begin{align}
	f''(r)=-n(n-1)a^{n/2} r^{n-2}.
\end{align}
Assuming $n>0$, we consider only the lowest order in $\sqrt{a} r$ for the right hand side of the field equation and find:
\begin{align}
	f(r)\frac{f(r)^2-1}{r^2}&=1-(\sqrt{a} r)^n)\frac{(\sqrt{a} r)^{2n}-2(\sqrt{a} r)^n}{r^2} \nonumber \\& =-2a^{n/2} r^{n-2}\quad \mathrm{as}\quad r\rightarrow 0.
\end{align}
Equating the above expressions implies that $n(n-1)=2$ which has the solutions $n=2,-1$. We assumed $n>0$ and thus only $n=2$ is a valid solution. Hence we conclude that the lowest order $n$ which solves the field equation \eqref{Field equation} is the integer solution $n=2$ that we already knew.

Now we do the corresponding considerations in the limit $r\rightarrow \infty$. From the energy formula:
\begin{align}
E=4\pi\int_0^\infty dr\ \left[f'^2+\frac{(f^2-1)^2}{2r^2}\right],
\end{align}
 it is clear that using \eqref{asymptoticrlarge} makes $f'(r)$ go like $ r^{-2(n+1)}$ for $r\rightarrow \infty$. This gives a convergent contribution to the energy in the limit $r\rightarrow \infty$ for all $n>0$. Using $n>0$ gives a convergent contribution for the second term in the energy integral as well. Thus we found the proper restriction on $n$ in \eqref{asymptoticrlarge}.
 
 Again we plug the asymptotic function \eqref{asymptoticrlarge} into the field equation \eqref{Field equation}:
\begin{align}
	n(n+1)\frac{b^{n}}{r^{n+1}}=\left(1-\frac{b^n}{r^{n}}\right)\frac{\left(1-\frac{b^{n}}{r^{n}}\right)^2-1}{r^2}.
\end{align}
Keeping only the lowest orders in $br^{-1}$ we get
\begin{align}
	n(n+1)\frac{b^n}{r^{n+2}}=2\frac{b^n}{r^{n+2}}.
\end{align}
Hence we obtain the equation $n(n+1)=2$ which has solutions $n=1$ and $n=-2$. And again we found no new information, as we already knew these solutions from the power series expansion \eqref{combined power series}.

\subsection{Non-existence of smooth singular finite energy solutions}

Now the results of the previous section is used to prove that not even singular smooth finite energy solutions exists of the form \eqref{Ansatz} except the trivial ones. For this we use the Non-existence theorem together with the asymptotic results of the previous subsection. We assume smoothness and show that it implies infinite energy.

Consider the  $r\rightarrow 0$ limit of the Wu-Yang gauge fields \eqref{Ansatz} with the asymptotic form of $f$ \eqref{asymptoticrsmall} and its parity transformation. As we saw in the previous subsection, both ensures a convergent energy contribution for small $r$. The Wu-Yang monopoles are
\begin{align}
A_i^a(\textbf{x})=-\epsilon_{iak}x^ka
\label{asymvacuum}
\end{align}
for $f(r)=1-ar^2$, and
\begin{align}
A_i^a(\textbf{x})=\epsilon_{iak}x^k\left(a-\frac{2}{r^2}\right)
\label{asymdiv}
\end{align}
for $f(r)=-1+ar^2$.

It is clear that a solution with the first behaviour \eqref{asymvacuum} has no singularities at $r=0$, also it is non-trivial and extended to all $r$ in a smooth manner. Hence the solution has infinite energy due to the Non-existence theorem. Notice now that the singular solution \eqref{asymdiv} has identical energy to \eqref{asymvacuum} since the energy formula \eqref{Energy} exhibits parity symmetry. Hence \eqref{asymdiv} has infinite energy. This proves that singular solutions have infinite energy because their non-singular parity transforms have so. 

This is an interesting case where the parity transform of a Wu-Yang monopole gives a different appearing monopole with resembling properties. 

\subsection{\label{sec:level8}A class of finite energy solutions}

The non-existence of smooth finite energy Wu-Yang monopoles is established. We are, however, able to construct finite energy Wu-Yang monopoles if we let go of the smoothness constraint. By allowing a single discontinuity in $f'$ the non-existence theorem \cite{coleman1977there} is no longer valid. This gives a class of solutions which fulfils \eqref{modified field equation} and may be of physical significance. They can be made both non-singular and with finite energy. Consider Marinho \textit{et al.} \cite{marinho2009revisiting} for several such solutions.

One way to construct finite energy solutions is to take a solution with non-diverging energy in the limit $r\rightarrow 0$ and make a cut-off when it moves outside the domain $[-1,1]$. The cut-off point is joined with the appropriate vacuum solution which makes the overall solution continuous. An example is seen in Figure \ref{fig2b} where we arbitrarily chose the parity $f(r)\rightarrow 1-ar^2$ in the limit $r\rightarrow 0$. This we call a cut-off solution. The only freedom in a cut-off solution is the length scale $a^{-1/2}$. Constructing cut-off solutions is a way of restricting the dynamics to a finite volume of space, thus avoiding the infrared $r\rightarrow\infty$ divergence of the energy formula \eqref{Energy}.

\begin{figure}
	\centering		
	\includegraphics[width=0.50\textwidth]{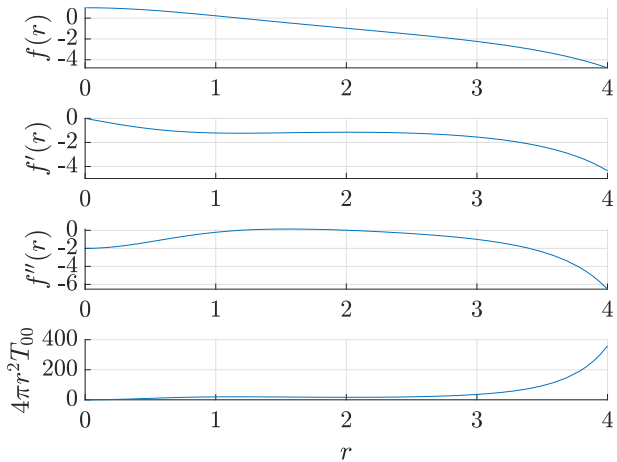}
	\caption{This figure shows the solution $f$ and the derivatives $f'$, $f''$, and the radial energy density $-4\pi r^2L(r)$, obtained from the boundary conditions $f(R)=1-R^2$ and $f'(R)=-2R$ with $R=0.01$. The radial axis is given in units of $a^{-1/2}$. All shown functions diverge when $r\rightarrow5.3$, and a beginning divergent behaviour is clearly seen in the energy density. Notice that $f=1-r^2$ when $r\ll 1$.}\label{fig2a}
\end{figure}

\begin{figure}
	\centering		
	\includegraphics[width=0.50\textwidth]{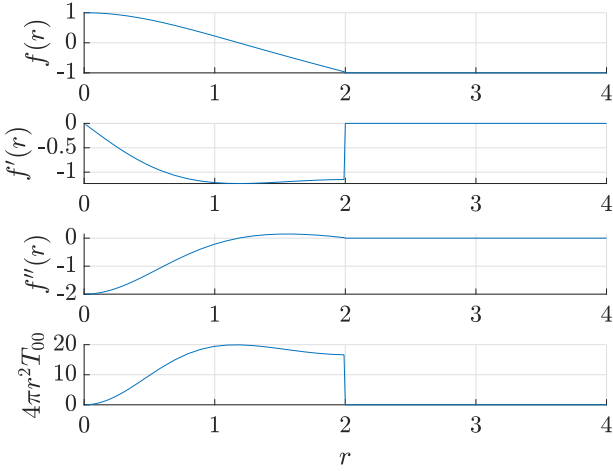}
	\caption{This figure shows the same functions as Figure \ref{fig2a} except that $f(r)=-1$ when $r\geq 2$. Notice the continuous behaviour of $f$, and the discontinuous behaviour of $f'$ and the radial energy density.}\label{fig2b}
\end{figure}

The construction assumes that all smooth solutions with finite energy in the limit $r\rightarrow 0$ moves outside $[-1,1]$ at some radius. We have not established this by analytical means. Nowhere did we prove that $f$ needs to venture outside $[-1,1]$ to gain infinite energy. However, this is a result which is strongly suggested by our numerical calculations. For the solution in Figure \ref{fig2a} we found the divergent behaviour of \eqref{cubed solution} at the radius $r=5.3a^{-1/2}$. The radial value where $f$ moves outside $[-1,1]$, we shall call the critical radius, $r_0$. The critical radius is unique. This is inferred from the proof of the Finite energy theorem (see Appendix \ref{Appendix3}). In other words, when a solution has left the domain $[-1,1]$, it is not going back.


In the solution of Figure \ref{fig2a}, we choose $a^{-1/2}$ as an arbitrary unit of length for the radial axis. It is seen from the graph that $\sqrt{a}r_0 = 2$ is fulfilled for the critical radius. This implies a fundamental relationship between the length scale of the problem and the critical radius. Such a relationship was expected on dimensional grounds, what is interesting is the value of the proportionality constant. Figure \ref{fig2a} suggests the constant of proportionality to be a factor of two. When constructing the graph, we assumed asymptotic behaviour at radius $R=0.01a^{-1/2}$ to fix the constants of integration. The estimated proportionality constant should converge towards the true value as we choose $R$ smaller. Calculations show that $R=0.01a^{-1/2}$ is reasonably small, since smaller $R$ results in $\sqrt{a} r_0=2$ as well.

Changing $a$ corresponds to doing a scale transformation. This is most easily seen in the small $\sqrt{a} r$ limit where a scale transformation $f(r)\rightarrow f(\lambda r)$ takes the form
\begin{align}
1- ar^2 \rightarrow 1- a\lambda^2 r^2.
\end{align}
Hence a scale transformation changes $a$ into $a\lambda^2$, which simply corresponds to redefining $a$.
Thus the solution in Figure \ref{fig2a} represents the equivalence class of all solutions with $f(r)\rightarrow 1- ar^2$ as $r\rightarrow 0$. In fact, all finite energy cut-off solutions may be constructed from the one in Figure \ref{fig2b} by varying $a$ and changing parity. Thus the whole set of finite energy cut-off solutions are classified by two equivalence classes represented by the solution in Figure \ref{fig2b} and its parity transformation.

One approach to finding the energy of a cut-off solution, is to calculate the area under the radial energy density graph of Figure \ref{fig2b}. This is numerically found to be 
\begin{align}
E=28\sqrt{a}=\frac{46}{r_0}.
\label{numericalenergy}
\end{align}
Thus the class of cut-off solutions are able to produce arbitrary energies by changing $a$, hence they produce an energy continuum. Equation \eqref{numericalenergy} has the feature that the smaller the system the larger the energy. This is expected on dimensional grounds. Physically, this may be understood in terms of work done on the system when the system is compressed into a smaller volume. Relating \eqref{numericalenergy} to the adiabatic first law of thermodynamics $dE=dW=-pdV$, we find that
\begin{align}
	dE=-\frac{46}{r_0^2}dr_0=-\frac{23}{2\pi r^4_0}dV,
	\label{first_law}
\end{align}
where we used $dV=4\pi r_0^2 dr_0$. Thus we find that the pressure exerted by the gauge fields is highly dependent on the critical radius, $p\propto r_0^{-4}$.

An analytical approach to finding the energy, is to use the usual energy formula \eqref{Energy} and take the critical radius $r_0$ as the upper integral limit:
\begin{align}
E=4\pi\int_0^{r_0} dr\ \left[f'^2+\frac{(f^2-1)^2}{2r^2}\right].
\end{align}
This seems reasonable since when $r>r_0$ the system is just a vacuum with vanishing energy density. Since $f$, per construction, has a converging energy in the $r\rightarrow 0$ limit, the above integral is finite. A crude analytical estimate of the proportionality constant between $E$ and $\sqrt{a}$ is made by using the asymptotic form of $f$ on the whole domain:
\begin{align}
	E&=4\pi\int_0^{r_0} dr\ \left[4a^2r^2+\frac{(a r^2-2\sqrt{a} r)^2}{2r^2}\right] \nonumber \\
	&=\frac{16 \pi}{3}a^2 r_0^3+ \frac{16 \pi}{3} a^2 r_0^3+8\pi a r_0-4\pi a^{3/2} r_0^2 \nonumber \\
	&=4\pi\left[\frac{8}{3}a^2 r_0^3-a^{3/2} r_0^2+2a r_0\right]= \frac{256 \pi}{3}\sqrt{a}.
\end{align}
Here we used $\sqrt{a}r_0=2$. It gives an overestimation by a factor of 10, and clearly the asymptotic approximation is no good except when $\sqrt{a} r\ll 1$. To obtain a better analytical estimate of the energy one may take more terms of the power series expansion \eqref{expansionA} into account.

We have to question the above discussion of the energy. One can argue that because the cut-off solutions fulfil the modified field equation \eqref{modified field equation} rather than the original one \eqref{Field equation}, the Lagrangian of the cut-off solutions should change and the energy as well. First we need a Lagrangian which reproduces the modified field equation. This is obtained from the old Lagrangian \eqref{Lagrangian} by simply subtracting the term $f(r)\zeta \delta(r-r_0)$,
\begin{align}
	L(r)=-\left(\frac{f'(r)}{r}\right)^2-\frac{(f(r)^2-1)^2}{2r^4}-f(r)\zeta\delta(r-r_0).
\end{align}
It is clear that $f(r)$ disappears doing the variation, leaving $\zeta\delta(r-r_0)$ alone. Thus we obtain \eqref{modified field equation} upon variation. The energy is found as before but with the new Lagrangian. Since
\begin{align}
		\zeta=f'(r_{0+})-f'(r_{0-})=-f'(r_{0-})
\end{align}
using \eqref{Zeta}, and $f'(r_{0+})=0$ for the vacuum, we have a new formula for the cut-off energy
\begin{align}
	E=4\pi\int_0^{r_0} dr\ \left[f'^2+\frac{(f^2-1)^2}{2r^2}\right]-4\pi f(r_0)f'(r_0).
	\label{modifiedEnergy}
\end{align}
For the class of solutions with parity as in Figure \ref{fig2b}, we have $f(r_0)=-1$ per construction, and we read of $f'(r_0)=-1.1\sqrt{a}$ from the graph. Doing a parity transformation, both values change sign and the energy remain unchanged. Thus the energy is still invariant under a parity transformation. All that is changed in the above analysis is the proportionality constant in \eqref{numericalenergy}. Now we have
\begin{align}
	E=\frac{46}{r_0}-\frac{8.8\pi}{r_0}=\frac{18}{r_0}=9\sqrt{a}.
	\label{modifiednumerical}
\end{align}
So the energy is more than halved by the introduction of the discontinuity. 

What kind of physical system may be modelled by the non-smooth cut-off Wu-Yang monopole? It is clear that the energy density drops to zero after the critical radius, as a consequence of construction. This gives the solution a glue-ball like appearance as appears of Figure \ref{fig2b}. Accordingly, the modelled system must be constraint in space, or at least approximately, to a ball with radius the critical radius. Another important feature of the cut-off solution is the inverse proportionality of the energy and size, which resembles a gas with pressure inversely proportional to $r_0^4$. Alternatively, a relation like \eqref{modifiednumerical} can be thought of as an uncertainty relation between position and energy of a point particle. 

Since the Wu-Yang monopole consists of three gauge fields, one might hope to model hadronic systems with three quarks, or any number of quarks. If we take a proton which has a charge radius of 0.8414 fm \cite{chargeradius}. Plugging this into the energy formula \eqref{modifiednumerical} gives
\begin{align}
	E_p=\frac{18}{0.84\mathrm{fm}}\cdot 0.20 \mathrm{GeV fm}=4.2 \mathrm{GeV}.
\end{align}
This energy is about five times to large to be the observed proton mass 0.9383 GeV. Thus the set of cut-off solutions does not give a proper model of the proton when identifying the charge radius with the critical radius. Had we used the peak energy density radius instead, the result would be better, but still of by more than a factor of two. The maximum energy density radius is read of Figure \ref{fig2b} to be $r_{max}=1.2\sqrt{a}$, then we get
\begin{align}
	E_p=\frac{11}{0.84\mathrm{fm}}\cdot 0.20 \mathrm{GeV fm}=2.6 \mathrm{GeV}.
\end{align}
In Marinho \textit{et al.} \cite{marinho2009revisiting} four discontinuous solutions are considered. They find that the discontinuity should be at a radius of at least four times the typical hadronic scale in order to obtain the proper masses. This agrees somewhat with our result for the proton.

The cut-off model allows for arbitrary critical radii, $r_0$, and thus one can add a time dependence on the problem, allowing $r_0$ to change. This gives rise to a change in energy \eqref{modifiednumerical} over time. When $r_0$ increases the system is doing work, and when $r_0$ decreases, work is done on the system. This act as a toy-model of a positively curved expanding universe with radius $r_0$. At present time, the expansion of the universe is dark energy dominated \cite{dodelson2003modern}, consistent with the dominating contribution of the universal energy density being constant. This is not realized by the cut-off model, which gives an average energy density of
\begin{align}
\rho_{universe}=\frac{18}{4\pi r_0^4}=\frac{1.4}{r_0^4}.
\end{align}
Rather, the cut-off model has the energy density of a radiation dominated universe, $\rho\propto a^{-4}$, where $a$ denotes the scale factor of the universe. Indeed the universe was radiation dominated at the earliest times. Since we know the average energy density of photons today from measurements of the Cosmic Microwave Background (CMB), we can estimate the radiation radius of the universe in the cut-off model. In other words, the cut-off model gives an estimate of the radius of a universe containing only radiation. The energy density of the CMB today is $0.25$ eV cm$^{-3}$ \cite{longair2013confrontation}. From this, we obtain the radiation radius of the universe:
\begin{align}
	r_0=\left(\frac{1.4\cdot 0.20\cdot 10^{-4}\, \mathrm{eV\, cm}}{0.25\, \mathrm{eV\, cm^{-3}}}\right)^{1/4}=0.10\, \mathrm{cm}.
\end{align}
This number is obviously to small to be the radius of our present day physical universe. However, it is the predicted radius of a radiation dominated universe. Since the universe was only radiation dominated at early times, we cannot compare the result to the universe we see around us. However, a quick estimate shows that the cut-off model must be off by several orders of magnitude. Already when the radiation dominated universe was $1$ s old, and the energy density was $10^6$ times larger, the universe must have been at least $3\cdot 10^8$ m in radius due to the speed of light. Since then, the universe has only grown larger. Thus the cut-off model seems not to be a realistic candidate of a cosmological model.

\section{\label{sec:level10}Generalizations of the Wu-Yang monopole}

We have been interested in generalizations of the Wu-Yang monopole to higher dimensional spaces, and have worked on different ideas to produce these. Unfortunately, as the reader will see, some particular roadblocks have been uncovered that have rendered the attempts thus far unsuccessful. In the end of the section we give an idea for a generalization which have not jet been made concrete. 

For the interested reader we give here some references that describe specific generalizations of the Wu-Yang monopole. A generalization of the Wu-Yang monopole to a solution with a topological charge is found in reference \cite{gogilidze1998wu}. Another form of the Wu-Yang monopole defined on the 2-unit sphere is found in reference \cite{hogan1983some}. Here it is further generalized to a solution of the $SU(n+1)$ Yang-Mills equations on the complex projective n-plane for $n\geq1$. In reference \cite{marciano1978magnetic} the Dirac monopole \cite{dirac1931quantised} is embedded into a $SU(2)$ theory in the abelian gauge. Here the Dirac string singularity is removed by a gauge transformation and the result is a natural generalization of the Wu-Yang monopole with $f=0$. The same idea of embedding Dirac monopoles into a non-abelian theory is used in reference \cite{popov2005explicit} to make a multi Wu-Yang monopole configuration with the individual monopoles placed at arbitrary points in space. t' Hooft considered the Wu-Yang monopole in the context of a gauge theory with a Higgs triplet \cite{t1974magnetic}. A generalization of the Wu-Yang monopole to a solution where 
\begin{align}
	A^a_0=x^a\frac{h(r)}{r^2},
\end{align}
was given by Julia and Zee (1975) in \cite{julia1975poles} in the context of a gauge theory with a Higgs triplet. Later Hsu and Mac (1977) considered the same solution in the context of a pure Yang-Mills theory \cite{hsu1977symmetry}, obtaining the field equations
\begin{align}
	&h''(r)=\frac{2h(r)f(r)^2}{r^2}, \\
	&f''(r)=f(r)\frac{f(r)^2-1-h(r)^2}{r^2},
\end{align}
which constitutes a natural generalization of the Wu-Yang field equation \eqref{Field equation}. This is further generalized to the case of a non-trivial topological solution with imaginary $A_0^a$ in \cite{wang2000static}. This gives a non-trivial non-singular finite energy solution.

In the quest for generalizations, we first ask what happens if another index is added in the Levi-Civita symbol. This is possible only if another space dimension and gauge field is allowed in the theory so that the indices can run consistently through 1,2,3,4. If we want to stay in a $SU(2)$ gauge theory, the extra gauge field needs to be removed. This is not the biggest concern, though. By using $\epsilon_{iajk}$ instead of $\epsilon_{iaj}$ one has an extra index which needs to be paired for the theory to make sense. The only possibility at hand is to use another coordinate $x^k$ giving
\begin{align}
	A^a_{i}=\epsilon_{iajk}x^jx^k\frac{g(r)}{r^2}.
\end{align}
From the antisymmetric properties of the Levi-Civita symbol we quickly realize that this anzats is identically zero for all the gauge fields.

This leads to the question whether another structure constant may be used instead of the Levi-Civita symbol. This is not possible since we need certain properties of the structure constant to do the calculations found in the appendix. These properties are anti-symmetry and the identity $\epsilon_{abc}\epsilon_{ijc}=\delta_{ai}\delta_{bj}-\delta_{aj}\delta_{bi}$. Since anti-symmetry is what needs to be avoided above, this approach is not going to work.

A promising approach of generalizing the Wu-Yang monopole to higher space dimensions is obtained from the results of Yang in reference \cite{yang1978generalization}. Here the Dirac monopole \cite{dirac1931quantised} is generalized to a $SU(2)$ monopole in a five dimensional space with a non-vanishing second Chern class number. As mentioned above, the generalization of a $f=0$ Wu-Yang monopole to a non-abelian Dirac monopole in three dimensional space is done in reference \cite{marciano1978magnetic}.

In Yang's work \cite{yang1978generalization}, we find the Lagrangian of the generalized Dirac monopole to have the simple form
\begin{align}
	L_{DM}(r)=-\frac{1}{4}F^a_{\mu\nu}F_a^{\mu\nu}=-\frac{3}{r^4}.
\end{align}
As we found earlier, the Wu-Yang Lagrangian is given by
\begin{align}
L_{WY}(r)=-\left(\frac{f'(r)}{r}\right)^2-\frac{(f(r)^2-1)^2}{2r^4}.
\label{L_WY}
\end{align}
We now ask if we can choose $f$ to make the Lagrangians appear identical. This requires that $f'=0$ hence $f$ must be constant. Choosing $f=0$ gives the wanted result up to a scale:
\begin{align}
	L_{WU}(r)=-\frac{1}{2r^4},
\end{align}
hence both Lagrangians result in the same field equations and the theories are identical in the $f=0$ limit. We expect that Yang's results \cite{yang1978generalization} can be generalized in some sense to include a general function $f$.

A first try to generalize the Yang monopole \cite{yang1978generalization}, is to allow five space dimensions in the Wu-Yang gauge fields \eqref{Ansatz}. This is not going to work, however, as argued above. We have not yet been able to work out a generalization which includes a function. Finding such a generalization would give a generalization of the Wu-Yang monopole to a five dimensional space.

\section{Conclusions and outlook}
Let us first summarize our main results, and then make some comments. In section \ref{sec:level1} we show that the gauge fields of the Wu-Yang monopole are not related by rotations, contrary to what might be suspected from their vector plots. We introduce a new way of thinking about the monopole charge $f(r)$ as a string in a centrifugal potential. The string picture gives a natural interpretation of the boundary conditions needed in the variation of the action. We consider Pad{\'e} approximant solutions of the field equation. It turns out that these simply reproduce the constant solutions. A novel form of the energy formula is derived using a Neumann boundary condition. This gives a relationship between the energetics of the oscillating and repulsive parts of a monopole charge.
We calculate the complicated chromomagnetic fields of the Wu-Yang monopole and the relevant charge densities. This shows that in the gauge dependent chromomagnetic field picture, the Wu-Yang solution is a collection of three dipoles. 

In Section \ref{sec:level5} we discus existence questions related to finite energy solutions. It is argued that no non-trivial smooth solutions exists of the Wu-Yang type. Not even singular ones. This is followed by a discussion of cut-off solutions, which form a particular class of numerical solutions which have a single discontinuity in the derivative. We find the implications of the discontinuity, and see that cut-off solutions are naturally thought of in terms of the first law of thermodynamics. The solution is applied as a model of the proton and a radiation dominated universe.

In future research, we encourage studying applications of non-smooth Wu-Yang monopoles. The simple cut-off solutions give rise to a rich array of possible applications. The possible existence of the five dimensional Yang monopole is particularly relevant at present time where Yang monopoles are being realized in the laboratory \cite{sugawa2016observation, yan2018yang}. Along the same line of reasoning, the Wu-Yang monopole might be realized in a laboratory as well. In the resent proposal \cite{jia2016analogous}, it is suggested that a Wu-Yang monopole may be simulated in a $^3$He superfluid system. In particular the realization of discontinues Wu-Yang monopoles is of interest as they appear to have a broad domain of theoretical application.

\appendix
\section{\label{Appendix}Appendix}

Here we give proofs of the results stated in the article. 

\subsection{\label{Coulomb gauge}Coulomb gauge}

In this appendix we show that the Wu-Yang monopole fulfils Coulomb gauge, $\partial^iA_i^a=0$. This is calculated as follows
\begin{align}
	\partial^i A_i^a&=\partial^i\left(\epsilon_{iaj}x^j\frac{g(r)}{r^2}\right)=\epsilon_{iaj}x^j\partial^i\left(\frac{g(r)}{r^2}\right) \nonumber \\ &=
	\epsilon_{iaj}x^j \frac{r^2\partial^ig(r)-g(r)\partial^i(r^2)}{r^4} \nonumber \\ &=
	\epsilon_{iaj}x^j \frac{rx^ig'(r)-2g(r)x^i}{r^4}=0,
\end{align}
since $\epsilon_{iaj}x^ix^j=0$ due to anti-symmetry. The second equality uses that for identical $i$ and $j$, $\epsilon_{iaj}=0$.

\subsection{\label{Appendix1}Field strengths}

We calculate the field strength components, $F^a_{ij}$, of the Wu-Yang monopole. We already know that $F^a_{0\nu}=0$ since the fields are time-independent and $A^a_0=0$. We find the space components $F^a_{ij}$ by the calculation
\begin{align}
F^a_{ij} &=\partial_iA^a_j-\partial_jA^a_i+\epsilon_{abc}A^b_iA^c_j \nonumber\\
&=[\partial_i(\epsilon _{jal} x^l)-
\partial_j(\epsilon_{iaj}x^j)]\frac{g}{r^2}+
\epsilon_{jal}x^l\partial_i\left(\frac{g}{r^2}\right)\nonumber\\&\quad -
\epsilon_{iaj}x^j \partial_j\left(\frac{g}{r^2}\right)
+\epsilon_{abc}\epsilon_{ibk}\epsilon_{jcl}x^kx^l \frac{g^2}{r^4}& \nonumber\\
&=[\epsilon_{jai}-\epsilon_{iaj}]\frac{g}{r^2}+[\epsilon_{jal}x^lx_i-\epsilon_{iak}x^kx_j]\left[\frac{g'(r)}{r^3}-2\frac{g}{r^4}\right] \nonumber\\
&\quad
+[\delta_{al}\delta_{bj}-\delta_{aj}\delta_{bl}]\epsilon_{ibk} x^kx^l \frac{g(r)^2}{r^4} \nonumber\\
&=2\epsilon_{ija}\frac{g}{r^2}+[\epsilon_{jal}x^lx_i-\epsilon_{ial}x^lx_j][\frac{g'}{r^3}-2\frac{g}{r^4}]\nonumber \\ & \quad+\epsilon_{ijl}x^lx^a\frac{g^2}{r^4}.
\end{align}
The third equality is obtained by the identity $\epsilon_{abc}\epsilon_{ljc}=\delta_{al}\delta_{bj}-\delta_{aj}\delta_{bl}$, and the fourth equality is obtained using the antisymmetry of $\epsilon_{abc}$. 

\subsection{\label{Appendix2}Wu-Yang Lagrangian}
Here we use the field strengths to calculate the Lagrangian of the Wu-Yang monopole. Throughout the calculation, we use the antisymmetry of $\epsilon_{abc}$ and the identities $x^ix_i=r^2$ and $\epsilon_{abc}\epsilon_{lbc}=2\delta_{al}$. We find that
\begin{align}
&F^a_{ik}F^{ik}_a= 4\epsilon_{ika}\epsilon_{ika}\frac{g^2}{r^4} \nonumber \\ &
+[\epsilon_{kal}x^lx_i-\epsilon_{iaj}x^jx_k][\epsilon_{kan}x^nx_i-\epsilon_{iam}x^mx_k]\left[\frac{g'}{r^3}-2\frac{g}{r^4}\right]^2
\nonumber\\&\nonumber
+\epsilon_{ikj}x^jx^a\epsilon_{ikl}x^lx^a\frac{g^4}{r^8} \\\nonumber&+
4\epsilon_{ika}[\epsilon_{kal}x^lx_i-\epsilon_{iaj}x^jx_k]\frac{g}{r^2}\left[\frac{g'}{r^3}-2\frac{g}{r^4}\right]\\& \nonumber+
4\epsilon_{ika}\epsilon_{ikj}x^jx^a\frac{g^3}{r^6}\\ \nonumber &+
2[\epsilon_{kal}x^lx_i-\epsilon_{iaj}x^jx_k]\epsilon_{ikj}x^jx^a\left[\frac{g'}{r^3}-2\frac{g}{r^4}\right]\frac{g^2}{r^4}\\&\nonumber
=24\frac{g^2}{r^4}+4x^lx_lx^ix_i\left[\frac{g'}{r^3}-2\frac{g}{r^4}\right]^2\\ \nonumber &+
2x^jx_jx^ax_a \frac{g^4}{r^8}+16x^lx_l\frac{g}{r^2}\left[\frac{g'}{r^3}-2\frac{g}{r^4}\right] +8x^ax_a\frac{g^3}{r^6}+0 \\\nonumber & 
=24\frac{g^2}{r^4}+4\left[\frac{g'}{r}-2\frac{g}{r^2}\right]^2+
2\frac{g^4}{r^4}+16g\left[\frac{g'}{r^3}-2\frac{g}{r^4}\right]+8\frac{g^3}{r^4}
\\&
=8\frac{g^2}{r^4}+4\frac{g'^2}{r^2}+2\frac{g^4}{r^4}+8\frac{g^3}{r^4}=4\left(\frac{g'}{r}\right)^2+4\frac{2g^2+2g^3+\frac{1}{2} g^4}{r^4}.
\end{align}
Hence the Lagrangian, $L=-\frac{1}{4}F^a_{ik}F_a^{ik}$, takes the form
\begin{align}
	L=-\left(\frac{g'}{r}\right)^2-\frac{2g^2+2g^3+\frac{1}{2} g^4}{r^4}.
\end{align}
By introducing $f:= g+1$ we may rewrite this like
\begin{align}
L=-\left(\frac{f'}{r}\right)^2-\frac{(f^2-1)^2}{2r^4}.
\end{align}
This is seen as follows
\begin{align}
	L&=-\left(\frac{f'}{r}\right)^2-\frac{(f^2-1)^2}{2r^4} \nonumber\\ &=-\left(\frac{[g+1]'}{r}\right)^2-\frac{\left([g+1]^2-1\right)^2}{2r^4} \nonumber \\&
	=-\left(\frac{g'}{r}\right)^2-\frac{\left(2g+g(r)^2\right)^2}{2r^4} \nonumber \\&
	=-\left(\frac{g'}{r}\right)^2-\frac{4g^2+g^4+4g^3}{2r^4}.
\end{align}

\subsection{\label{Recursion formula} Power series solution}

Assuming a solution of the field equation
\begin{align}
f''(r)=f(r)\frac{f(r)^2-1}{r^2}
\end{align} 
has a power series expansion at $r=0$, we consider
\begin{align}
A(r)=\sum_{n=0}^{\infty}a_n r^n.
\end{align}
Plugging this into the field equation gives
\begin{align}
\sum_{n=2}^{\infty}n(n&-1)a_n r^{n-2} \nonumber \\
&=\sum_{l,m,n=0}^{\infty}a_l a_m a_n r^{l+m+n-2}-\sum_{n=0}^\infty a_n r^{n-2}.
\label{coefficients}
\end{align}

Due to the first term on the right hand side, the coefficients follow a recursion relation. However, this is not a closed recursion relation. It becomes increasingly more complex for higher order coefficients since the higher order, the more possibilities of combining $l,m,n$ to give the correct order in the exponent of $r^{l+m+n-2}$.

The left hand side has no negative powers of $r$ implying that $a_0^3-a_0=0$ and $3a_1a_0^2-a_1=0$ do to linear independence of the set $\{r^n:n\in \mathbb{Z}\}$. The first equality implies that $a_0=0,\pm1$ which implies that $a_1=0$. Including the contributions from the left hand side we find with $a_1=0$ three more coefficients,
\begin{align}
&2a_2=3a_0^2 a_2+3a_0a_1^2-a_2 \nonumber \\
&\quad\;\; =3a_0^2 a_2-a_2, \\
&6a_3=3a_3a_0^2+a_1^3+3a_2 a_1 a_0-a_3\nonumber\\
&\quad\;\;=3a_3a_0^2-a_3, \\
&12a_4=3a_4 a_0^2+3a_3a_1a_0+3a_2^2a_0+3a_2a_1^2-a_4 \nonumber \\
&\quad\quad =3a_4 a_0^2+3a_2^2a_0-a_4.
\end{align}
Thus if $a_0^2=1$ we obtain
\begin{align}
a_2=a_2,\qquad a_3=0, \qquad a_4=\frac{3}{10}a_0a_2^2
\end{align}
It is clear that the complexity of the coefficients increases with the number af combinations that realizes the order of the given coefficient. It is clear that if $a_0=0$, all these coefficients vanish and therefore all higher coefficients must as well giving the trivial solution.  When $a_0^2= 1$ we see that $a_3=0$. This must be true for all higher order uneven coefficients as well since all terms in the first sum on the right-hand side of \eqref{coefficients} must contain uneven coefficients to add up to an uneven order. Taking $a_0^2=1$, the relevant expression looks like
\begin{align}
n(n-1)a_n=2a_n+\sum \mathrm{terms\,containing\,uneven\,a_{i<n}},
\end{align}
where the last part disappears because all lower uneven coefficients vanish. Then because $n>2$ we get $a_n=0$ for uneven $n$.

It is rather interesting to notice that $a_4$ picks up a factor of $a_0$. As we explain in Subsection \ref{solutions}, this is true of all coefficients having $a_2$ to an even power.

For $a_0=\pm 1$, the coefficient $a_2$ is the only free parameter. It has units of inverse length squared which in natural units is energy squared. Hence the free parameter $\sqrt{a_2}$ sets the energy scale. This is clear from the cut-off solution \eqref{modifiednumerical}

Assuming a solution has a power series expansion at $r=\infty$, we consider
\begin{align}
B(r)=\sum_{n=0}^{\infty}b_n r^{-n},
\end{align}
and get the following field equation
\begin{align}
\sum_{n=0}^{\infty}n(n &+1)b_n r^{-(n+2)} \nonumber\\ 
&=\sum_{l,m,n=0}^{\infty}\left[b_l b_m b_n r^{-(l+m+n+2)}-b_n r^{-(n+2)}\right].
\end{align}
Notice that this field equation is different from the corresponding one for $A(r)$. One might have expected to obtain a symmetry when doing the transformation $l,m,n\rightarrow-l,-m,-n$ (leaving the summation unchanged), however, this is not the case as seen from the left hand side which starts at $n=0$, unlike the corresponding equation for $A(r)$. This means that $b_1$ is generally not zero which implies that the uneven coefficients are generally not zero. Thus the corresponding recursion formula for the $b_n$ coefficients is more involved than that of the $a_n$ coefficients.

The zeroth coefficient again fulfils the equation $b_0^3-b_0=0$ which has solutions $b_0=0,\pm1$. The first order coefficient fulfils $2b_1=3b_1 b_0^2-b_1$ which makes $b_1$ arbitrary when $b_0=\pm1$. Therefore $b_1$ is a free parameter of length dimensions which sets the energy scale, $b_1^{-1}$, of the solution. When $b_0=0$ then also $b_1=0$ and so, like before, all other coefficients vanish, resulting in the trivial constant solution.

\subsection{\label{Pade}Pad{\'e} approximant solution}

Here we use a Pad{\'e} approximant ansatz for the field equation \eqref{Field equation}. The ansatz is as follows:
\begin{align}
	f(r)=\frac{\sum_{m=0}^{M}a_mr^m}{1+\sum_{n=1}^{N}b_nr^n}.
\end{align}
We shall take $N=M$ for simplicity, and write $b_0=1$ to allow compact notation. The left-hand side of the field equation \eqref{Field equation} becomes
\begin{align}
	&f''(r)=\mathrm{\frac{d}{dr}}\left[\frac{\sum_{m,n=0}^N (m-n)a_mb_nr^{m+n-1}
	}{\sum_{n,m=0}^{N}b_nb_mr^{n+m}}\right] \nonumber \\
&=\frac{\sum_{k,l,m,n=0}^N(m-n)(m+n-k-l-1)a_mb_nb_kb_lr^{k+l+m+n-2}}
{\sum_{k,l,m,n=0}^N b_kb_lb_mb_nr^{k+l+m+n}}.
\end{align}
The right-hand side of the field equation becomes
\begin{align}
&\frac{f(r)^3-f(r)}{r^2}=\frac{\sum_{l,m,n=0}^{N} a_l(a_ma_n-b_mb_n)r^{l+m+n-2}}{\sum_{l,m,n=0}^{N} b_lb_mb_nr^{l+m+n}} \nonumber \\
&=\frac{\sum_{k,l,m,n=0}^{N} b_ka_l(a_ma_n-b_mb_n)r^{k+l+m+n-2}}{\sum_{k,l,m,n=0}^{N} b_kb_lb_mb_nr^{k+l+m+n}}.
\end{align}
Equating the two expressions using the common denominator, we obtain a relationship among coefficients for each exponent, $K-2$, of $r$: 
\begin{align}
	\sum_{k+l+m+n=K}[(m-n)&(m+n-k-l-1)a_mb_nb_kb_l-\nonumber \\ 
	& b_ka_l(a_ma_n-b_mb_n)]=0
	\label{Padecoef}
\end{align}
This is the same method we used for the power series expansion. The zero order coefficients are found from the $k=l=m=n=0$ terms. Using $b_0=1$, we obtain
\begin{align}
	0=a_0(a_0^2-1).
\end{align}
This equation has the solutions $a_0=0,\pm1$, which is exactly the same result we found for the zeroth order coefficients of the power series expansions in Appendix \ref{Recursion formula}. Thus the power series solutions and the Pad{\'e} approximant solution agree when $r\approx0$. Choosing $a_0=0$ gives the trivial solution $f=0$, so we choose to consider only $a_1=\pm1$. When $k+l+m+n=1$ there are $4$ possibilities of combining the indices. We obtain
\begin{align}
	0=2(a_1-a_0b_1),
\end{align}
implying that $a_1=a_0b_1$. Thus truncating the Pad{\'e} approximant at $N=2$ gives the constant solutions, $f=\pm1$ when $a_0=\pm1$.

When $k+l+m+n=2$ there are $10$ possibilities of combining the indices. The following equation is obtained
\begin{align}
	-3a_1b_1-a_0a_1b_1+5a_0b_1^2+b_1^2-2a_0a_1^2=0.
\end{align}
This equations contains no information of $a_2$ and $b_2$, and turns out to be trivial, $0=0$, when $a_1=a_0b_1$ is used. Thus we need higher order terms than $r^0$ to determine the constraints on $a_2$ and $b_2$. To do this calculation we consider the specific Pad{\'e} approximant
\begin{align}
	f(r)=\frac{a_0+ar^2}{1+br^2},
\end{align}
where we have written $a_2=a$ and $b_2=b$ for notational simplicity and chosen $a_1=0=b_1$ to simplify further. Using that $a_0^2=1$, we obtain the left-hand side of the field equation 
\begin{align}
	f''(r)=\frac{2(a-a_0b)(1+6br^2+5b^2r^4)}{(1+br^2)^4},
\end{align}
and the right-hand side of the field equation
\begin{align}
	&\frac{f(r)^3-f(r)}{r^2}=\frac{2(a-a_0b)+3a_0(a^2-b^2)r^2}{(1+br^2)^4}+ \nonumber \\ &\frac{\left([a^3-a_0b^3]+3ab[aa_0-b]\right)r^4+ab(a^2-b^2)r^6}{(1+br^2)^4}.
\end{align}
We require equality, and see that the $r^0$ equation is indeed trivial $2(a-a_0b)=2(a-a_0b)$. This is consistent with our result using the general equation \eqref{Padecoef}. Going to higher exponents in $r$, it is clear that there is a contribution to $r^6$ only on the right-hand side. Thus we obtain $0=ab(a^2-b^2)$ which has the non-trivial solution $a^2=b^2$. We find the $r^2$ and $r^4$ equations to be consistent if $a=a_0b$. Thus again we obtain equality between the coefficients up to a sign, $a_2=a_0b_2$.

For the exponents where $k+l+m+n=3$, there are $20$ possible combinations of the coefficients. Choosing for notational simplicity $a_0=1$ so that $a_1=b_1$, we obtain
\begin{align}
	&6a_3-6b_3-2a_1b_2+2b_2b_1-2a_1b_2+b_2b_1+2a_2b_1\nonumber \\&-2b_2a_1-2a_1b_1^2+2b_1^3=2(a_3-b_3)+6(a_2-b_2)a_1.
\end{align}
Which reduce to
\begin{align}
	a_3-b_3=(a_2-b_2)b=0.
\end{align}
Thus we see again that $a_3=b_3$, and a general calculation shows that $a_3=a_0b_3$. Assuming this trend continues, we have shown that Pad{\'e} approximants gives no new solutions for the Wu-Yang field equation. They give the constant solutions $f=0,\pm 1$.
\subsection{\label{Magnetic field}Magnetic fields and charge densities}

In this appendix we calculate the magnetic fields, $B^a_i=\frac{1}{2} \epsilon_{ijk}F^a_{jk}$, of the Wu-Yang monopole. We use the identity $\epsilon_{abc}\epsilon_{aij}=\delta_{bi}\delta_{cj}-\delta_{bj}\delta_{ci}$ to get rid of the Levi-Civita symbols.

\begin{align}
		B^a_i&= \epsilon_{ijk}[\epsilon_{kal}x^lx_j-\epsilon_{jal}x^lx_k]\left[\frac{g'(r)}{2r^3}-\frac{g(r)}{r^4}\right]\nonumber \\&\quad+\epsilon_{ijk}\epsilon_{jkl}x^lx^a\frac{g(r)^2}{2r^4}+\epsilon_{ijk}	
	\epsilon_{jka}\frac{g(r)}{r^2} 
	\nonumber \\&
	=[(\delta_{ia}\delta_{jl}-\delta_{il}\delta_{ja})x^lx_j\nonumber \\ &\quad-(\delta_{ka}\delta_{il}-\delta_{kl}\delta_{ia})x^lx_k]\left[\frac{g'(r)}{2r^3}-\frac{g(r)}{r^4}\right]
	\nonumber \\&\quad+2\delta_{il}x^lx^a\frac{g(r)^2}{2r^4}+2\delta_{ia}\frac{g(r)}{r^2} 
	\nonumber \\&
	=[\delta_{ia}x^jx_j-x^ix_a]\left[\frac{g'(r)}{r^3}-2\frac{g(r)}{r^4}\right]
	\nonumber \\&\quad+x_ix^a\frac{g(r)^2}{r^4}+2\delta_{ia}\frac{g(r)}{r^2} 
\end{align}

To find the associated magnetic charge densities, we now calculate the divergence of the magnetic fields
\begin{align}
	\partial_i B^a_i&=\partial_a\frac{g'(r)}{r}-2\partial_a\frac{g(r)}{r^2}-\partial_i \left(x_ix_a\frac{g'(r)}{r^3}\right) \nonumber \\
	&\quad+2\partial_i\left(x_ix_a\frac{g(r)}{r^4}\right)+\partial_i\left(x_ix_a\frac{g(r)^2}{r^4}\right)+2\partial_a\frac{g(r)}{r^2} \nonumber \\ 
	&=\frac{g''(r)}{r^2}x_a-\frac{g'(r)}{r^3}x_a-2\frac{g'(r)}{r^3}x_a+4\frac{g(r)}{r^4}x_a \nonumber \\
	&\quad-3x_a\frac{g'(r)}{r^3}-x_i\delta_{ia}\frac{g'(r)}{r^3}-x_ix_a\frac{g''(r)}{r^4}x_i \nonumber \\
	&\quad+3x_ix_a\frac{g'(r)}{r^5}x_i+6x_a\frac{g(r)}{r^4}+2x_i\delta_{ia}\frac{g(r)}{r^4} \nonumber \\
	&\quad+2x_ix_a\frac{g'(r)}{r^5}x_i-8x_ix_a\frac{g(r)}{r^6}x_i+3x_a\frac{g(r)^2}{r^4} \nonumber \\
	&\quad +x_i\delta_{ia}\frac{g(r)^2}{r^4}+2x_ix_a\frac{g(r)g'(r)}{r^5}x_i \nonumber \\
	&\quad-4x_ix_a\frac{g(r)^2}{r^6}x_i+2\frac{g'(r)}{r^3}x_a-4\frac{g(r)}{r^4}x_a \nonumber \\
	&=2x_a\frac{g(r)g'(r)}{r^3}.
\end{align}
All but one term cancel, giving the final equality.
\subsection{\label{Appendix3}Finite energy theorem}

Here we give a proof of the result argued in Marinho \textit{et al.} \cite{marinho2009revisiting}, that in order for the Wu-Yang monopole to possess finite energy,
\begin{equation}
E=4\pi\int_0^\infty dr \left\{f'^2+\frac{(f^2-1)^2}{2r^2}\right\}<\infty,
\end{equation}
$f$ must lie in the interval $[-1,1]$ for all $r$. We assume only that $f$ is two times differentiable. This is natural since it fulfils a second order differential equation. Notice that this assumption assures that $f''$ is continuous do to the field equation \eqref{Field equation}.

The argument goes by showing the contrapositive statement: if there is a point $r$ where $f$ is outside $[-1,1]$, then the energy becomes infinite. There are several cases and we do them separately.

Assume first that $f$ increases, not necessarily monotonically, from some value within the interval $[-1,1]$ to some value larger than one. Then obviously some $r_0$ has the property that $f(r_0)>1$, and due to continuity there is some $a$ with the property that $f(a)=1$. In order for $f$ to reach $f(r_0)$ from below there must be some $\tilde{r}$ fulfilling $a<\tilde{r}<r_0$ where $f'(\tilde{r})>0$ and $f(\tilde{r})>1$ because $\tilde{r}>a$. Due to the field equation, we also have $f''(\tilde{r})>0$. But then $f'$ is increasing on some small interval around $\tilde{r}$, and so is $f$ because $f'(\tilde{r})>0$. But $f''$ is increasing when $f$ increases due to the field equation, and so there is a self-perpetuating effect making $f$ and $f'$ increase for all $r>\tilde{r}$. But then $f'\nrightarrow0$ when $r\rightarrow\infty$, and the energy becomes infinite due to the $f'^2$ term. 

Assume now that $f(\tilde{r})>1$ for some $\tilde{r}\geq0$, and assume without loss of generality that $f'(\tilde{r})<0$; otherwise the situation is described by the argument above. If $\tilde{r}=0$, then due to continuity $f>1$ on some interval $[0,r_0]$, and the energy becomes infinite due to the $r^{-2}$ term. If $\tilde{r}>0$ then we may assume that $f'<0$ and $f>1$ on the interval $[0,\tilde{r}]$, and the energy diverges due to $r^{-2}$. Otherwise due to continuity, there has to be some $r_0$ in $[0,\tilde{r}]$ where $f'(r_0)\geq0$ and $f(r_0)>1$. But then $f''(r_0)>0$ and so there is some $r_1>r_0$ where $f'(r_1)>0$ and $f(r_1)>0$ and the situation is like the first part of the proof. 

For cases where $f(r)<-1$ for some $r$, consider $-f$ in the above arguments and note that the energy depends only on $f^2$. This completes the proof.

\subsection{\label{Appendix4}Energy continuum theorem}

An immediate consequence of the scaling invariance of the field equation \eqref{Field equation} is that one can construct solutions of arbitrary energy from just one finite energy solution.

Assume $f$ is a finite energy solution, then the solution $h(r)=f(\lambda r)$ with $\lambda>0$ has energy
\begin{align}
E[h]&=4\pi\int_0^{\infty} dr \left\{h'^2(r)+\frac{(h^2(r)-1)^2}{2r^2}\right\}\nonumber \\ &=4\pi\int_0^{\infty} dr \left\{\lambda^2f'^2(\rho)+\lambda^2\frac{(f^2(\rho)-1)^2}{2\rho^2}\right\}\nonumber\\&
=
4\pi\int_0^{\infty} d\rho \left\{\lambda f'^2(\rho)+\lambda\frac{(f^2(\rho)-1)^2}{2\rho^2}\right\}=\lambda E[f],
\end{align}
writing $\rho=\lambda r$. Since $\lambda$ is an arbitrary positive number, $E[h]$ may be chosen arbitrarily. 



\bibliography{references}
\end{document}